\documentclass[12pt]{iopart}

\usepackage{iopams}  
\usepackage{pdfpages}
\usepackage{amstext} 
\usepackage{graphicx}                       
\usepackage{overpic}
\usepackage{array}  
\usepackage{hyperref} 
\hypersetup{colorlinks   = true,linkcolor    = blue}

\begin{document}
\setlength{\parskip}{0pt}


\includepdf[pages=1]{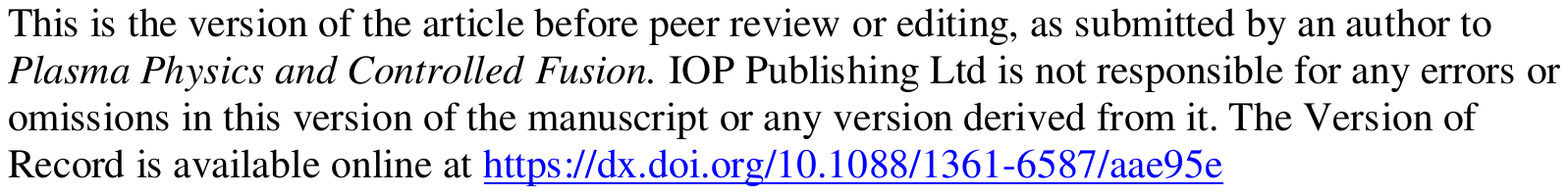}

\title[X-B conversion in density fluctuations]{Mode-conversion of the extraordinary wave at the upper hybrid resonance in the presence of small-amplitude density fluctuations}

\author{N A Lopez$^1$ and A K Ram$^2$}
\address{$^1$ Department of Astrophysical Sciences, Princeton University, Princeton, New Jersey 08544, USA}
\address{$^2$ Plasma Science and Fusion Center, Massachusetts Institute of Technology, Cambridge, Massachusetts 02139, USA}

\ead{nlopez@pppl.gov}

\begin{abstract}
In spherical tokamaks, the electron plasma frequency is greater than the electron cyclotron frequency. Electromagnetic waves in the electron cyclotron range of frequencies are unsuitable for directly heating such plasmas due to their reduced accessibility. However, mode-conversion of the extraordinary wave to the electron Bernstein wave (X-B mode-conversion) at the upper hybrid resonance makes it possible to efficiently couple externally-launched electromagnetic wave energy into an overdense plasma core. Traditional mode-conversion models describe an X-mode wave propagating in a potential containing two cutoffs that bracket a single wave resonance. Often, however, the mode-conversion region is in the edge, where turbulent fluctuations and blobs can generate abrupt cutoffs and scattering of the incident X-mode wave.  We present a new framework for studying the X-B mode-conversion which makes the inclusion of these fluctuations analytically tractable. In the new approach, the high-field cutoff is modelled as an infinite barrier, which manifests as a boundary condition applied to a wave equation involving only one cutoff adjacent to the resonance on the low-field side. The new model reproduces the main features of the previous approach, yet is more suitable for analyzing experimental observations and extrapolating to higher dimensions. We then develop an analytical estimate for the effect of small-amplitude, quasi-monochromatic density fluctuations on the X-B mode-conversion efficiency using perturbation theory. We find that Bragg backscattering of the launched X-mode wave reduces the mode-conversion efficiency significantly when the fluctuation wavenumber is resonant with the wavenumber of the incident X-mode wave. These analytical results are corroborated by numerically integrating the mode-conversion equations.

\end{abstract}

%
%
%
%
%

\section{Introduction}

Electron cyclotron (EC) waves are often used as an external means to heat plasmas to thermonuclear temperatures\cite{Ott80}, and to non-inductively drive toroidal current in tokamak systems\cite{Fisch80,Bornatici83,Erckmann94,Lloyd98,Prater04}. Indeed, EC waves are projected to play a key role in the achievement of fully non-solenoidal tokamak operation in NSTX-U\cite{Poli15}. In recent years, spherical tokamaks\cite{Peng86,Jardin03} (STs) have become increasingly prominent, due in large part to their compact size and attractive stability properties. However, since STs typically operate in an overdense regime, it is difficult to use traditional EC waves on STs for heating and current drive purposes.

Fortunately, the electron Bernstein wave (EBW) provides a method to heat an overdense plasma in the EC frequency range. The EBW\cite{Bernstein58,Laqua07} is a thermal mode which exists in the vicinity of the upper hybrid resonance (UHR) and the harmonics of the electron cyclotron resonance. It freely propagates within discrete frequency bands\cite{Crawford65,Puri73}, which can be roughly mapped into discrete spatial bands for an inhomogeneous plasma. Importantly, the EBW which originates at the UHR propagates unimpeded towards larger magnetic field until it reaches the nearest cyclotron resonance, where it will damp strongly\cite{Montes86,Ram00,Decker06}. This feature makes the EBW ideal for heating overdense plasmas.

Being a thermal mode of predominantly electrostatic polarization, the EBW does not propagate in vacuum; the excitation of the EBW via external means is non-trivial. Grills are not typically used due to the small grid spacing required to excite a wave whose wavelength is comparable to the electron gyroradius\cite{Laqua07}. Instead, experiments often use finely-tuned mode-conversions to excite the EBW from vacuum-launched electromagnetic (EM) EC waves. These EM-EBW mode-conversions can be decomposed into two basic categories: those that use a vacuum-launched ordinary mode (O-mode) wave to excite the EBW, and those that use a vacuum-launched extraordinary mode (X-mode) wave to excite the EBW.

The first category of EM-EBW mode-conversions, known as the O-X-B mode-conversion, relies on the coalescence of the O-mode and X-mode dispersion curves at oblique angles of propagation with respect to the vacuum magnetic field\cite{Preinhaelter73,Hansen85}. An O-mode wave injected at the critical angle will enter the coalescence region and mode-convert to the slow X-mode. The slow X-mode will then propagate to the UHR and excite the EBW\cite{Stix65}. The O-X-B mode-conversion has been extensively studied theoretically\cite{Mjolhus84,Gospodchikov06,Popov07,Popov10,Popov11}, and has been demonstrated experimentally on a number of devices\cite{Laqua03,Shevchenko07,Pochelon07,Yoshimura17}. This mode-conversion is best utilized when the characteristic length scales of the plasma equilibrium are large\cite{Ram03,Igami06}, in which case the O-X-B mode-conversion is largely decoupled from the second category of EM-EBW mode-conversions.

This second category of EM-EBW mode-conversions, known as the X-B mode-conversion, relies on an evanescent mode-coupling between the fast X-mode, which propagates only in low-density regions, and the slow X-mode, which propagates only in high-density regions\cite{Ram00,Piliya05}. The slow X-mode will then reach the UHR and excite the EBW. Like the O-X-B mode-conversion, the X-B mode-conversion has been well-demonstrated experimentally\cite{Shiraiwa06,Uchijima15,Seltzman17}; unlike the O-X-B mode-conversion, however, the X-B mode-conversion is difficult to study analytically, and is typically studied computationally with particle-in-cell methods\cite{Asgarian14,Xiao15,Arefiev17} or full-wave solvers\cite{Kim14,Jo17}. This trend can probably be attributed in large part to the inherent non-locality of the X-B mode-conversion, which will be elucidated in the following section. In contrast, the O-X-B mode-conversion is often very localized in physical space, which makes for straightforward analytics via Taylor expansions.

In this work, we present a new analytical and computational paradigm, which we title the `Infinite Barrier' model, to understand the X-B mode-conversion problem: in its simplest state, the X-B mode-conversion is nothing more than a particular solution to the Budden problem, which describes the electromagnetic wave propagation through an isolated cutoff-resonance pair\cite{Budden61,Budden79}. Given a model equation describing the Budden problem in a chosen equilibrium configuration, a solution to the X-B problem is then obtained by imposing a homogeneous (Dirichlet-type) boundary condition of the form $\textbf{y}(\textbf{L}) = 0$ to the Budden model equation, where $\textbf{L}$ parameterizes the location of the high-field side (HFS) X-mode cutoff. Within this framework the X-B problem should be easy to study numerically, as it becomes a straightforward exercise in the integration of boundary-valued partial-differential equations on a finite domain. An important difference between the new approach and previous models\cite{Ram96} is that the exponentially-growing X-mode field solution is not required to be completely absent within the mode-conversion region; this oft-employed criterion can be the source of large numerical error if the growing solution is erroneously excited via inexact arithmetic. By changing the boundary condition, the new model avoids this issue entirely.

For STs in particular, the X-B mode-conversion typically occurs in the edge plasma region, where fluctuations in local plasma parameters can deteriorate the mode-conversion efficiency\cite{Jones03,Taylor03}. In other contexts, density fluctuations are known to modify the wavenumber spectrum, and to produce spurious reflections of an incident wave field\cite{Ram13,Ram16}. Using the new Infinite Barrier model, we analytically investigate the effect of small-amplitude density fluctuations on the X-B mode-conversion efficiency. We find that resonant reflection of the launched X-mode wave can significantly inhibit the mode-conversion to the EBW at the UHR. This effect may explain some of the discrepancies between experimental observations and the existing theory for the X-B mode-conversion efficiency\cite{Shiraiwa06}.

This paper is organized as follows: Section \ref{1D} applies the new X-B mode-conversion model to a one-dimensional (1-D) inhomogeneous plasma and compares the result with an existing model. Section \ref{fluctuation} presents the main result of this work, the impact of small-amplitude density fluctuations on the X-B conversion efficiency. Finally, we conclude with a summary, and some comments on possible extensions of this work. Although our envisioned application for this approach is computational, the analysis presented in this paper is largely analytical as a proof of principle. Additionally, the analysis is restricted to 1-D to make an analytical description possible; for numerical applications, the Infinite Barrier model is immediately generalizable to higher dimensions.

\section{The X-B mode-conversion in 1-D}
\label{1D}

Before elaborating on the new model paradigm, let us first review some basic aspects of the X-B mode-conversion physics in 1-D. Consider an X-mode wave propagating in a plane-stratified inhomogeneous, stationary plasma, with RF-induced fluctuations having a time-dependence of the form $e^{-i\omega t}$. When the direction of inhomogeneity aligns with the direction of propagation, the y-component of the X-mode wave electric field satisfies the following differential equation\cite{Ram00,White74}:
\begin{equation}
E_y''(x) + Q(x) E_y(x) = 0
\label{Xwave}
\end{equation}

\noindent where generically:
\begin{equation}
Q(x) = \frac{S^2(x) - D^2(x)}{S(x)}
\label{potential}
\end{equation}

\noindent with $S(x)$ and $D(x)$ being the Stix functions, given in the high-frequency limit as\cite{Stix92}:
\numparts
\begin{eqnarray}
S(x) = 1 - \frac{\omega_{pe}^2(x)}{\omega^2-\omega_{ce}^2(x)}\\
D(x) = -\frac{\omega_{ce}(x)}{\omega}\frac{\omega_{pe}^2(x)}{\omega^2-\omega_{ce}^2(x)}
\end{eqnarray}
\endnumparts

\noindent Here, $\omega$, $\omega_{ce}$, $\omega_{pe}$ are the X-mode wave frequency, the electron cyclotron frequency, and the electron plasma frequency, respectively. The coordinate system is chosen such that $\hat{x}$ is aligned with the direction of inhomogeneity, $\hat{z}$ is aligned with the external magnetic field, and $\hat{y}$ is mutually orthogonal to both $\hat{x}$ and $\hat{z}$ in a right-handed sense. In writing equation \ref{Xwave}, all coordinates have been non-dimensionalized via the factor $\frac{\omega}{c}$, where $c$ is the speed of light.

\begin{figure}
\begin{overpic}[width=0.45\linewidth]{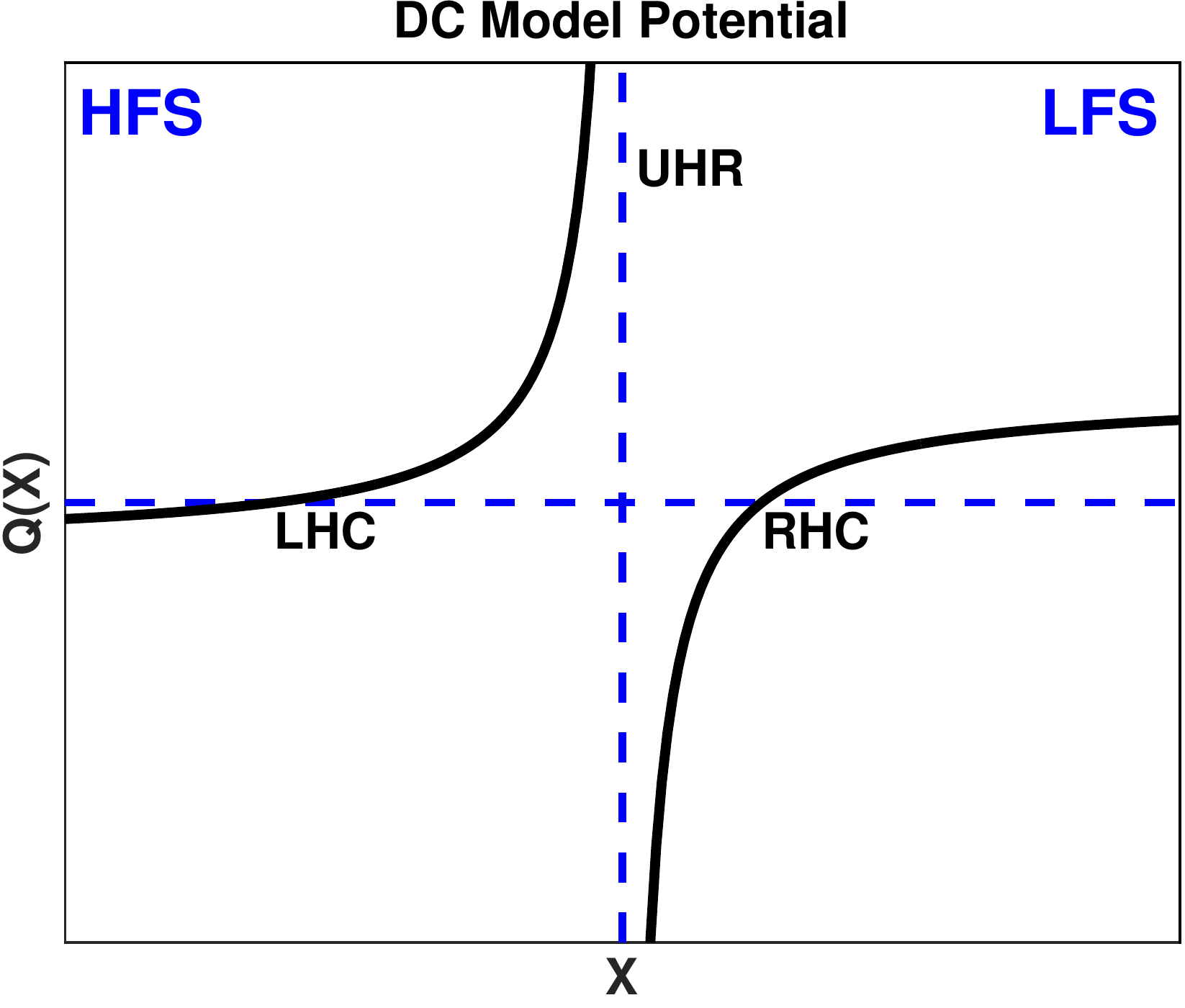}
\put(8,10){\large\textbf{(a)}}
\end{overpic}
\hspace{3mm}\begin{overpic}[width=0.45\linewidth]{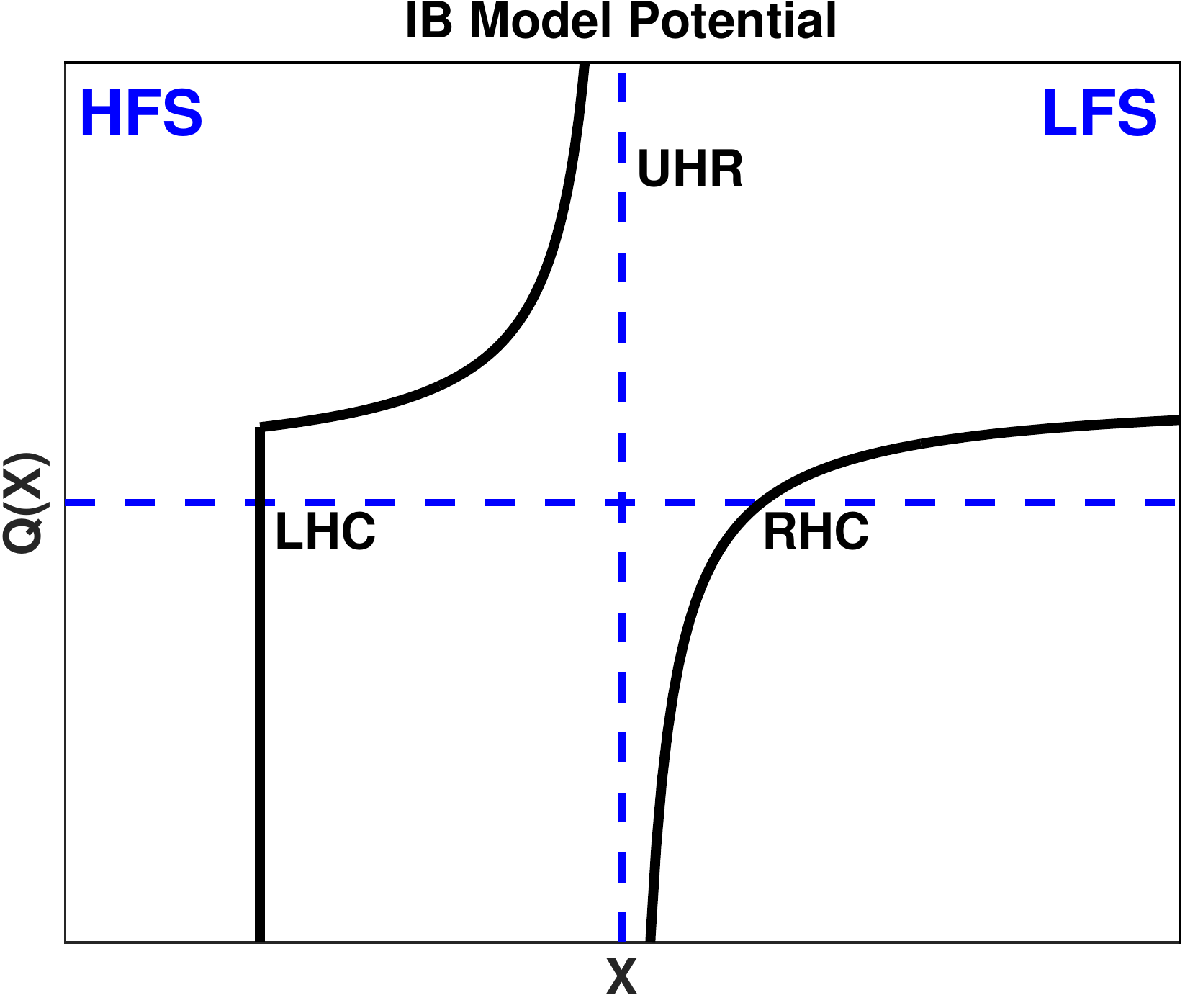}
\put(8,10){\large\textbf{(b)}}
\label{1D}
\end{overpic}
\caption{Two choices of model potential function $Q(x)$ for the X-B mode-conversion in 1-D. \textbf{(a)} `Double-cutoff' model, which generates the cutoff-resonance-cutoff triplet with two Budden potentials. \textbf{(b)} `Infinite Barrier' model, which generates the cutoff-resonance-cutoff triplet with a Budden potential and an infinite barrier.}
\label{1Dmodel}
\end{figure}

When propagating into increasing density and magnetic fields, the X-mode dispersion relation typically exhibits a cutoff-resonance-cutoff triplet structure, composed of the low-field right-hand cutoff (RHC), the intermediate-field UHR, and the high-field left-hand cutoff (LHC), as seen in figure \ref{1Dmodel}. The spatial regions that lie between the RHC and the UHR, and beyond the LHC are regions of wave evanescence. The establishment of such a cutoff-resonance-cutoff triplet defines the X-B mode-conversion scheme\cite{Ram00,Ram96}.

To obtain the steady-state response, we will ultimately model the X-B mode-conversion as a boundary-value problem. However, it is conceptually easier to understand the physics by appealing to a time-dependent description. In this picture, an X-mode wave incident from the low-field side (LFS) will impinge upon the RHC and the corresponding region of evanescence. A fraction of the wave energy will successfully tunnel beyond this region of evanescence into the propagating region bounded by the UHR and the LHC. Some of the tunnelled wave energy will be mode-converted to the EBW at the UHR, while the remaining tunnelled wave energy will propagate towards the LHC, reflect therein, and return to the UHR. In the steady-state, an interference pattern in the mode-conversion efficiency is thus established, akin to a standing wave in an optical cavity\cite{Ram96,Jackson75}. Notably, it becomes possible to obtain complete mode-conversion within the cutoff-resonance-cutoff triplet. This result is not obtainable unless the presence of the LHC is accounted for in the analysis.

\begin{figure}
\includegraphics[scale=0.8]{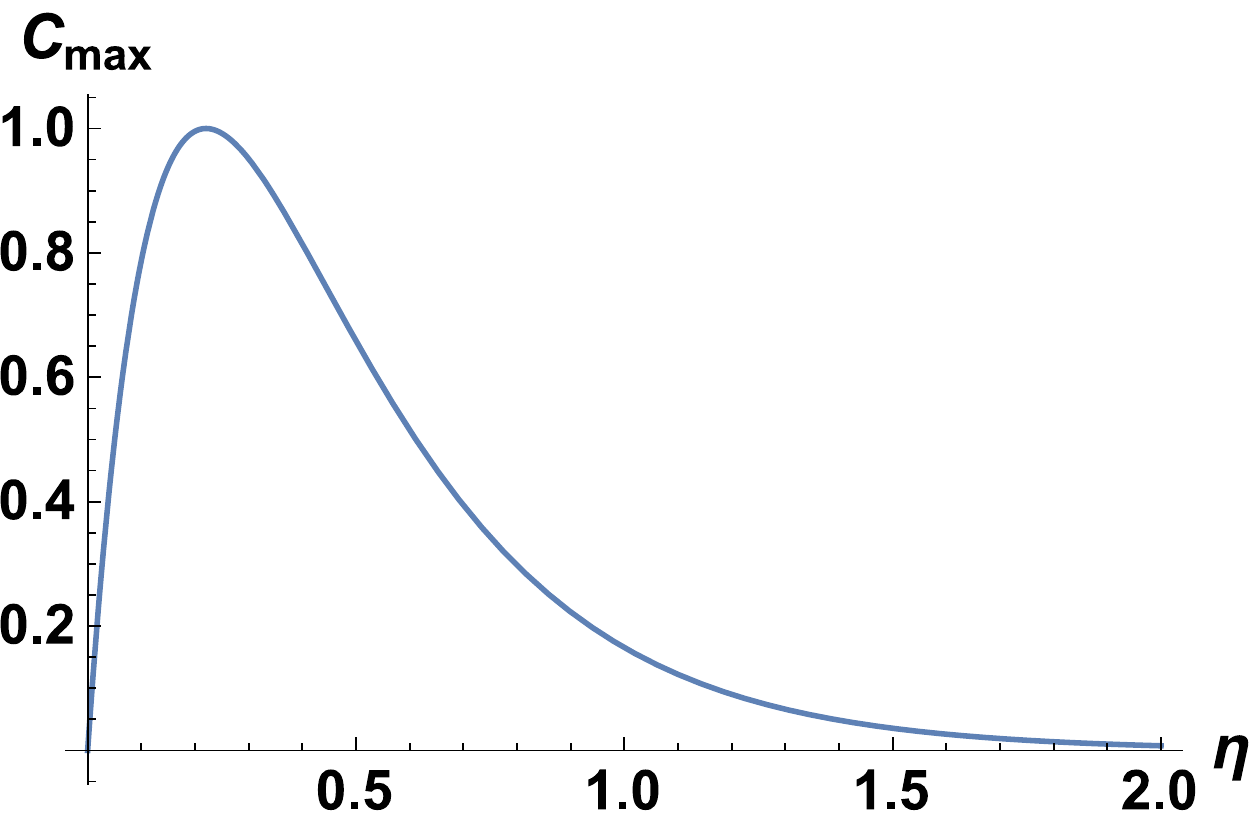}
\caption{The maximum X-B mode-conversion efficiency as a function of the Budden tunnelling parameter $\eta$, which is the width of the evanescent region normalized to the launched wavenumber of the incident X-mode wave. This maximum is attained when $\cos^2\left(\phi \right) = 1$, and there is constructive interference within the cutoff-resonance-cutoff triplet.}
\label{envel}
\end{figure}

As a general result, the X-B mode-conversion efficiency is given by the equation\cite{Ram96}:
\begin{equation}
C_\text{XB} = 4e^{-\pi\eta}\left(1-e^{-\pi\eta} \right)\cos^2\left(\phi \right)
\label{convEFF}
\end{equation}

\noindent where $\eta$ is sometimes known as the Budden tunnelling parameter. This is because the Budden problem, that is, the transmission of an X-mode wave through an isolated cutoff-resonance pair, only depends on the single parameter $\eta$\cite{Budden61}. Physically, $\eta$ represents the distance the X-mode must tunnel through between the RHC and the UHR, normalized by the launched wavelength. The specific form for the interference phase $\phi$ is highly model-dependent, and will be the focus of the remainder of section \ref{1D}. The envelope term to equation \ref{convEFF}, $C_\text{max} \doteq 4e^{-\pi\eta}\left(1-e^{-\pi\eta} \right)$, is shown in figure \ref{envel}. One can see that perfect mode-conversion is possible when $\eta = \frac{\log 2}{\pi}\approx 0.22$ and $\phi$ is a multiple of $\pi$.

\subsection{Double-cutoff model}

The Double-cutoff (DC) model is an exactly-solvable model for the interference phase $\phi$, presented originally in \cite{Ram96}. This model uses the potential function:
\begin{equation}
\label{DCmodel}
Q_\text{DC}(x) = \left\{ \begin{array}{l l}
\gamma_R - \frac{\beta}{x} & x > 0\\
-\kappa_L - \frac{\beta}{x} & x \le 0
\end{array}\right.
\end{equation}

\noindent where $\beta$, $\gamma_R$, and $\kappa_L$ are all positive constants. If $\kappa_L$ were negative, then this potential would correspond to the Budden potential, which contains just a resonance-cutoff pair.

In writing equation \ref{DCmodel}, we choose our coordinate origin such that the UHR is located at $x=0$. The normalized distance between the UHR and the RHC can be expressed as:
\begin{equation}
\sqrt{\gamma_R} d = \eta_R
\label{dDC}
\end{equation}

\noindent while the normalized distance between the UHR and the LHC can be expressed as:
\begin{equation}
\sqrt{\kappa_L} L = \eta_L
\label{LDC}
\end{equation}

\noindent where $\eta_R \doteq \frac{\beta}{\sqrt{\gamma_R}}$ and $\eta_L \doteq \frac{\beta}{\sqrt{\kappa_L}}$. Matching the solution across the pole $x=0$ is performed subject to the continuity conditions:
\numparts
\begin{eqnarray}
\label{mCond1}
E_y(0^+) &= E_y(0^-)\\
\label{mCond2}
E_y'(0^+) &= E_y'(0^-) - i\pi \beta y(0)
\end{eqnarray}
\endnumparts

\noindent Implicit in the above matching conditions is the choice of branch cut, such that $\log(0^+) = \log(0^-) - i\pi$. Physically, this choice of branch cut ensures causality - it is the analytic continuation of a damped solution when collisional dissipation is included into the X-mode dispersion relation.

Let us denote the solution in the regions $x > 0$, $x\le 0$ by $y_R(x)$, $y_L(x)$ respectively. With proper variable transformations, it can be shown that the Budden differential equation reduces to Whittaker's confluent differential equation; both $y_R(x)$ and $y_L(x)$ are therefore expressible as linear combinations of the Whittaker functions \cite{Budden61,Whittaker03}. Specifically, let us expand these solutions as:
\numparts
\begin{eqnarray}
\label{DCfielda}
y_R(x) = A \cdot W_{-i\frac{\eta_R}{2},\frac{1}{2}}\left(2\sqrt{\gamma_R}xe^{-i\frac{\pi}{2}}\right) + B \cdot W_{i\frac{\eta_R}{2},\frac{1}{2}}\left(2\sqrt{\gamma_R}xe^{-i\frac{3\pi}{2}}\right)\\
y_L(x) = C \cdot W_{-\frac{\eta_L}{2},\frac{1}{2}}\left(2\sqrt{\kappa_L}x\right) + D \cdot W_{\frac{\eta_L}{2},\frac{1}{2}}\left(2\sqrt{\kappa_L}xe^{-i\pi}\right)
\label{DCfieldb}
\end{eqnarray}
\endnumparts

\noindent Here, we have been explicit in the choice of Riemann sheet upon which the Whittaker functions should be evaluated. In the following, we shall suppress the second parameter of the Whittaker function; it should be understood that $W_k(x) \doteq W_{k,\frac{1}{2}}(x)$. 

There are four undetermined coefficients in equations \ref{DCfielda} and \ref{DCfieldb}; however, only three ratios need to be specified via boundary conditions: $\frac{A}{B}$, $\frac{C}{B}$, and $\frac{D}{B}$. With these three ratios specified, the coefficient $B$ provides an unimportant overall scaling to the wave fields.

The first boundary condition to be applied to the DC model is that the solution be subdominant on the real line for $x\to-\infty$. This `radiation' boundary condition requires that $C=0$. The other two boundary conditions for the DC model are obtained by imposing the matching conditions given in equations \ref{mCond1} and \ref{mCond2}, using the limits of the Whittaker functions as $x\to 0$ provided in \ref{WhitAsym}. As shown in \ref{PHIder}, the reflection coefficient for the DC model is obtained as:
\begin{equation}
R =1-4e^{-\pi\eta}\left(1-e^{-\pi\eta} \right)\cos^2\left(\phi_\text{DC} \right)
\end{equation}

\noindent with the phase $\phi_\text{DC}$ given as:
\begin{eqnarray}
\label{DCphase}
\phi_\text{DC} =\tan^{-1}\left(\frac{\log\left(\frac{\eta_R}{\eta_L}\right)+\Psi\left( 1-\frac{\eta_L}{2}\right)-\text{Re}\left[\Psi\left( 1-i\frac{\eta_R}{2}\right)\right]}{\text{Im}\left[\Psi\left( 1-i\frac{\eta_R}{2}\right)\right]-\frac{\pi}{2}} \right)
\end{eqnarray}

The presence of the HFS cutoff means that there can be no X-mode wave energy transmitted beyond the cutoff-resonance-cutoff triplet. By conservation of energy, this implies that the mode-conversion coefficient of the X-mode to the EBW at the UHR can be determined through the relation:
\begin{equation}
C_\text{XB} = 1 - R = 4e^{-\pi\eta}\left(1-e^{-\pi\eta} \right)\cos^2\left(\phi_\text{DC} \right)
\end{equation}

\noindent Hence, we confirm that the overall mode-conversion efficiency is indeed given by equation \ref{convEFF} for the DC model. This technique of identifying a mode-conversion coefficient with an amplitude sink of a reduced full-wave differential equation is sometimes known as the `resonant absorption' model of mode-conversion\cite{LashmoreDavies88,Cally10}.

\begin{figure}
\includegraphics[width=0.5\linewidth]{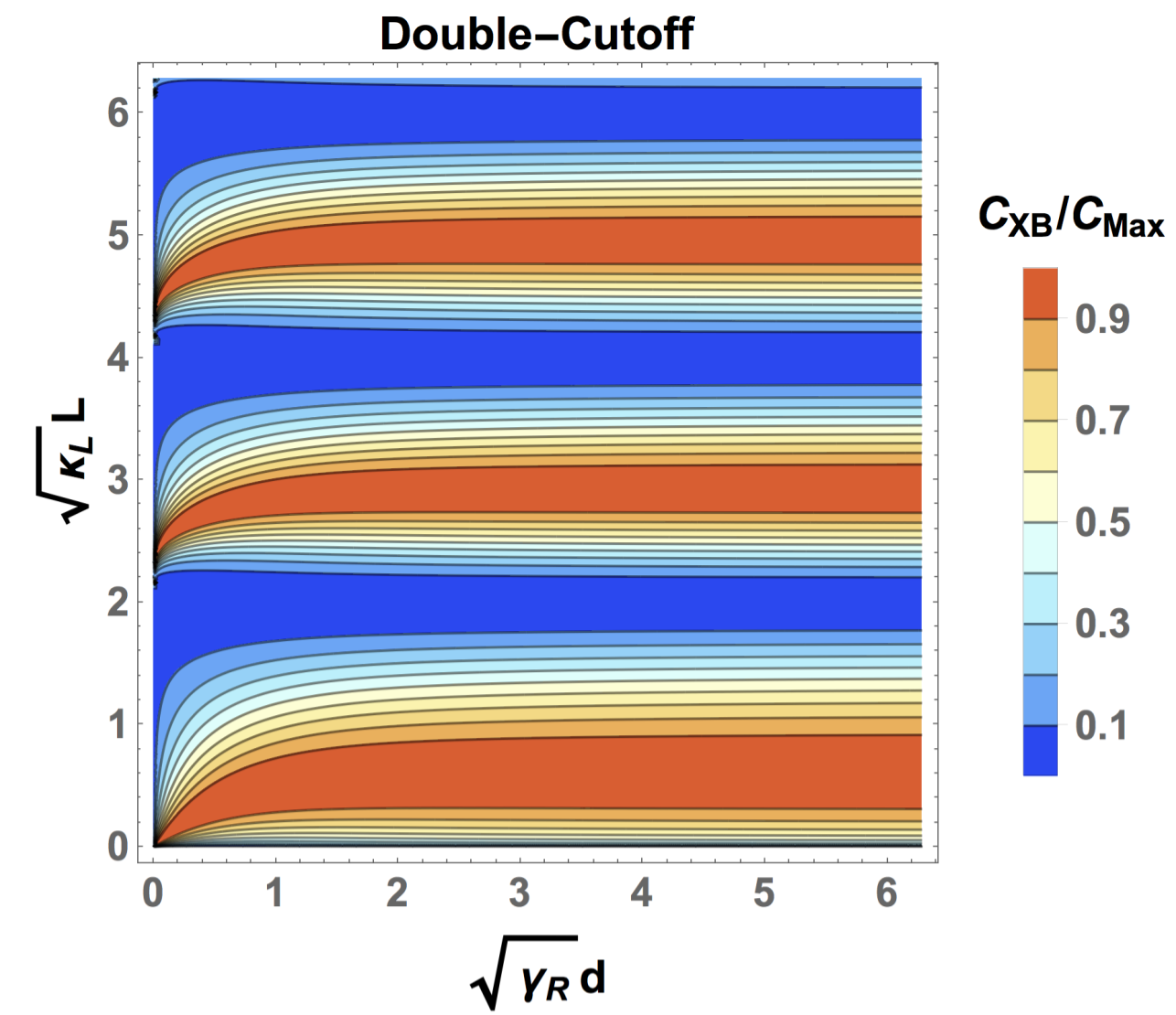}
\caption{Contour plot showing the variation of the oscillatory factor $\cos^2\left(\phi_\text{DC} \right)$ for the Double-cutoff model of the X-B mode-conversion efficiency as the distance between the right-hand cutoff and the upper hybrid resonance ($d$), and the distance between the upper hybrid resonance and the left-hand cutoff ($L$) are varied. When $L$ and $d$ are specified, the Double-cutoff model is completely determined.}
\label{DC_CONT}
\end{figure}

Figure \ref{DC_CONT} shows the variation of the oscillatory factor $\cos^2\left(\phi_\text{DC} \right)$ for the DC model as the locations of the high-density and low-density cutoffs are varied with respect to the UHR. For convenience, the distance between the UHR and the RHC is normalized by the launched wavenumber as $\sqrt{\gamma_R}d$, while the distance between the UHR and the LHC is normalized by the asymptotic decay coefficient of the HFS wave as $\sqrt{\kappa_L}L$. From the figure, one sees that the interference phase of the DC model is much more sensitive to variations in $\sqrt{\kappa_L}L$ than in $\sqrt{\gamma_R}d$: the phase exhibits a quasi-periodicity with respect to variations in $\sqrt{\kappa_L}L$ due to the wave interference within the cutoff-resonance-cutoff triplet cavity, while the phase is nearly independent of variation in $\sqrt{\gamma_R}d$ for $\sqrt{\gamma_R}d \gtrsim 1$. Importantly, the DC model successfully recovers the result $C_\text{XB} = 0$ for the confluent case $L=0$, when $Q(x)$ reduces to the Airy potential and the resonance disappears. Note that the functional form of the envelope $C_\text{max}$, shown in figure \ref{envel}, ensures that $C_\text{XB} = 0$ for the other confluent case $d=0$, irrespective of the phase function.

The strong dependence of $\phi_\text{DC}$ on $L$ has important consequences for X-B experiments on STs. For typical ST operating parameters, the RHC, the UHR, and the LHC are each located within the plasma edge region. The plasma edge region is often marred with large-scale density fluctuations born from turbulence and blobs. Dedicated EBW emission experiments on CDX-U and NSTX observed that the X-B coupling was complicated significantly by density fluctuations on the order of $80\%$ relative magnitude within the mode-conversion region\cite{Jones03,Taylor03}. Such density fluctuations can produce very abrupt cutoffs in the X-mode dispersion relation. This is problematic for the DC model, as one would need to measure the density profile with incredible resolution to compute $L$ as accurately as the high sensitivity of $\phi_\text{DC}$ would require. An alternative approach is to build the abrupt behavior of the LHC directly into the model, which would relax the demands on the measurement resolution. This approach defines the Infinite Barrier model, which is discussed in the following section.

\subsection{Infinite Barrier model in 1-D}

The DC model constructs the cutoff-resonance-cutoff triplet potential by analytically continuing the Budden problem to the situation when the asymptotic wavenumber of the HFS wave is imaginary. A simpler, more intuitive approach is offered by the Infinite Barrier (IB) model. It is well-established that the addition of infinite barriers, that is, regions in physical space where the field solution to a wave equation is identically zero, generates solution behavior that mimics classical particles reflecting off a solid interface\cite{Jackson75}. In a textbook example from quantum mechanics, infinite barriers are used to confine particles to a box\cite{Shankar94}. Using this logic, the IB model produces the cutoff-resonance-cutoff triplet by appending an infinite barrier to the Budden potential at the location of the HFS cutoff (see figure \ref{1Dmodel}). The infinite barrier produces the HFS cutoff, the Budden potential produces the remaining resonance-LFS cutoff pair. 

In practice, the IB model simply requires solving for the particular solution to the Budden problem with the conducting boundary condition $E_y(-L) = 0$, for some separation distance $L$ between the LHC and the UHR. This formulates the calculation of the X-B mode-conversion efficiency as a boundary-value problem on a finite domain. The finite domain is spanned by the launcher location at the LFS, and the HFS cutoff. This formulation is particularly amenable to computational implementation, since realistically, all computers solve differential equations on a finite domain. In contrast, by using a radiation boundary condition, the DC model is defined on an infinite domain. There are well-documented errors associated with domain-truncation when attempting to implement a radiation boundary condition on a computer\cite{Israeli81}. An example is the reflection of the wave field solution off the domain edge, which will introduce erroneous interference patterns into the X-B mode-conversion efficiency. These domain-truncation errors are avoided with the IB model.

The IB model uses the following choice of potential function in equation \ref{Xwave}:
\begin{equation}
Q_\text{IB}(x) = \left\{ \begin{array}{l l}
\gamma_R - \frac{\beta}{x} & x > 0\\[1mm]
\gamma_L - \frac{\beta}{x} & -L \le x \le 0\\[1mm]
-\infty & x < -L
\end{array}\right.
\end{equation}

\noindent where $\gamma_R$, $\gamma_L$, $\beta$, and $L$ are all positive constants. As with the DC model, we orient our coordinate origin such that the UHR is located at $x=0$. Like the DC model, the normalized distance between the UHR and the RHC is given as:
\begin{equation}
\sqrt{\gamma_R} d = \eta
\label{dIB}
\end{equation}

\noindent where $\eta \doteq \frac{\beta}{\sqrt{\gamma_R}}$. Unlike the DC model, the normalized distance between the UHR and the LHC, $\sqrt{\gamma_L}L$, is specified through the location of the infinite barrier, $L$.

As with the DC model, the solution within the region $x\in [-L,\infty)$ are linear combinations of Whittaker functions. Letting $y_R(x)$ be the field solution on the domain $x\in (0,\infty)$ and $y_L(x)$ be the field solution of the domain $x\in [-L,0]$, the general solution can be expressed as:
\numparts
\begin{eqnarray}
y_R(x) = A \cdot W_{-i\frac{\eta}{2}}\left(2\sqrt{\gamma_R}xe^{-i\frac{\pi}{2}}\right) + B \cdot W_{i\frac{\eta}{2}}\left(2\sqrt{\gamma_R}xe^{-i\frac{3\pi}{2}}\right)\\
y_L(x) = C \cdot W_{-i\frac{\eta r}{2}}\left(2\frac{\sqrt{\gamma_R}}{r}xe^{-i\frac{\pi}{2}}\right) + D \cdot W_{i\frac{\eta r}{2}}\left(2\frac{\sqrt{\gamma_R}}{r}xe^{-i\frac{3\pi}{2}}\right)
\end{eqnarray}
\endnumparts

\noindent Here, we have defined the asymmetry ratio $r \doteq \sqrt{\frac{\gamma_R}{\gamma_L}}$.

The analysis of the IB model is similar to that of the DC model; an important difference is the replacement of the radiation boundary condition, $C=0$, with the IB boundary condition, $y_L(-L) = 0$, to ensure regularity of the solution on the domain $x < -L$. As shown in \ref{PHIder}, the IB model recovers the general form for the conversion efficiency provided by equation \ref{convEFF}, with the interference phase given as:
\begin{eqnarray}
\hspace{-2.6cm}\phi_\text{IB} = \tan^{-1}\left(\frac{\sin\left(2\theta_L + 2\theta_W \right)\left[\pi + \pi\coth\left(\pi\frac{\eta r}{2} \right) -\Upsilon_\text{I} - \frac{2}{\eta r}\right]+\Upsilon_\text{R}\left[\cos\left(2\theta_L + 2\theta_W \right)+1\right]}{\cos\left(2\theta_L + 2\theta_W \right)\left[\pi + \pi\coth\left(\pi\frac{\eta r}{2} \right) -\Upsilon_\text{I} - \frac{2}{\eta r}\right]-\Upsilon_\text{R}\sin\left(2\theta_L + 2\theta_W \right) - \Upsilon_\text{I}} \right)\nonumber\\
\hspace{-1.5cm}- \theta_L - \theta_W 
\label{IBphase}
\end{eqnarray}

\begin{figure}
\begin{overpic}[width=0.5\linewidth]{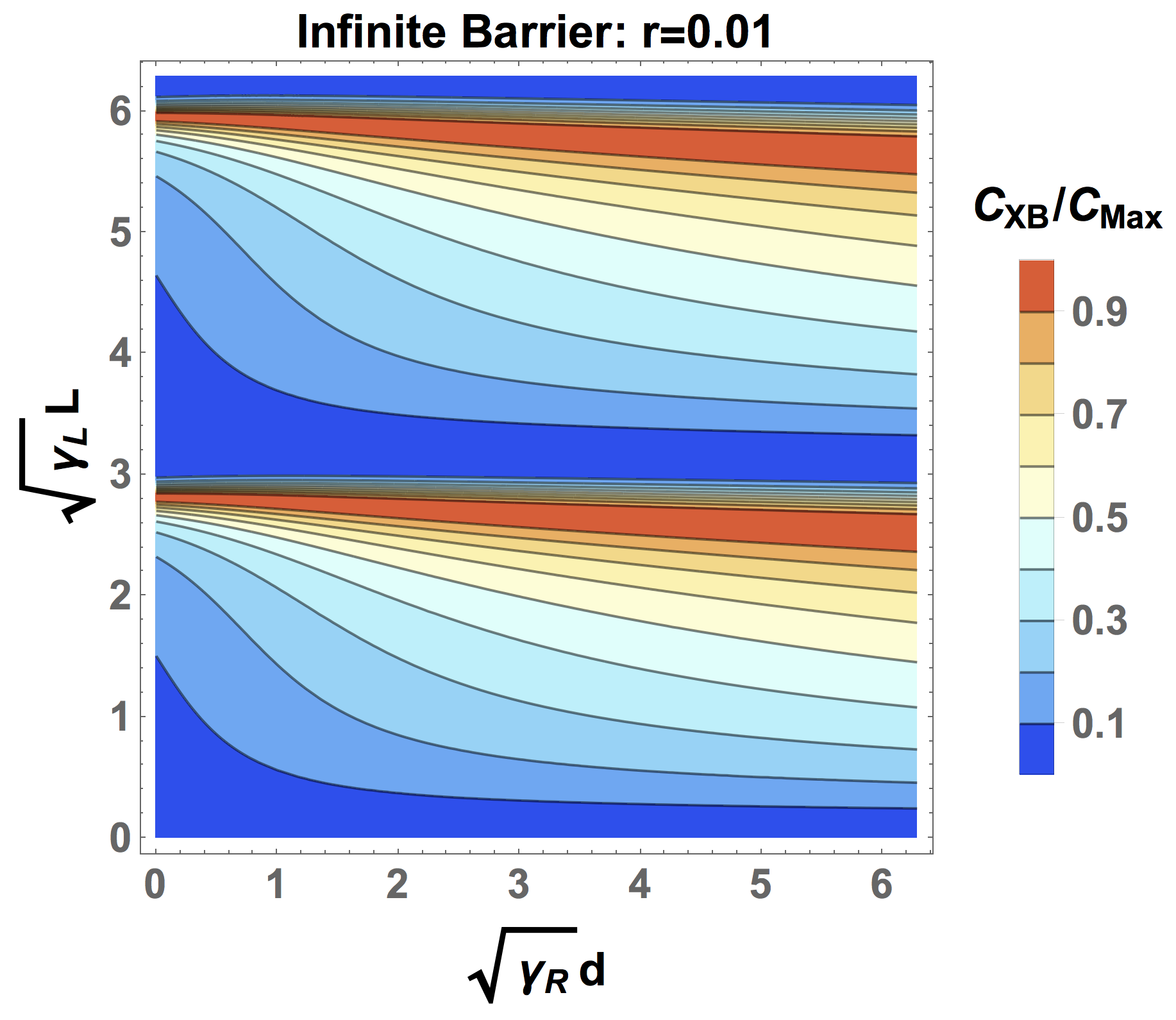}
\put(0,18){\large\textbf{(a)}}
\end{overpic}
\begin{overpic}[width=0.5\linewidth]{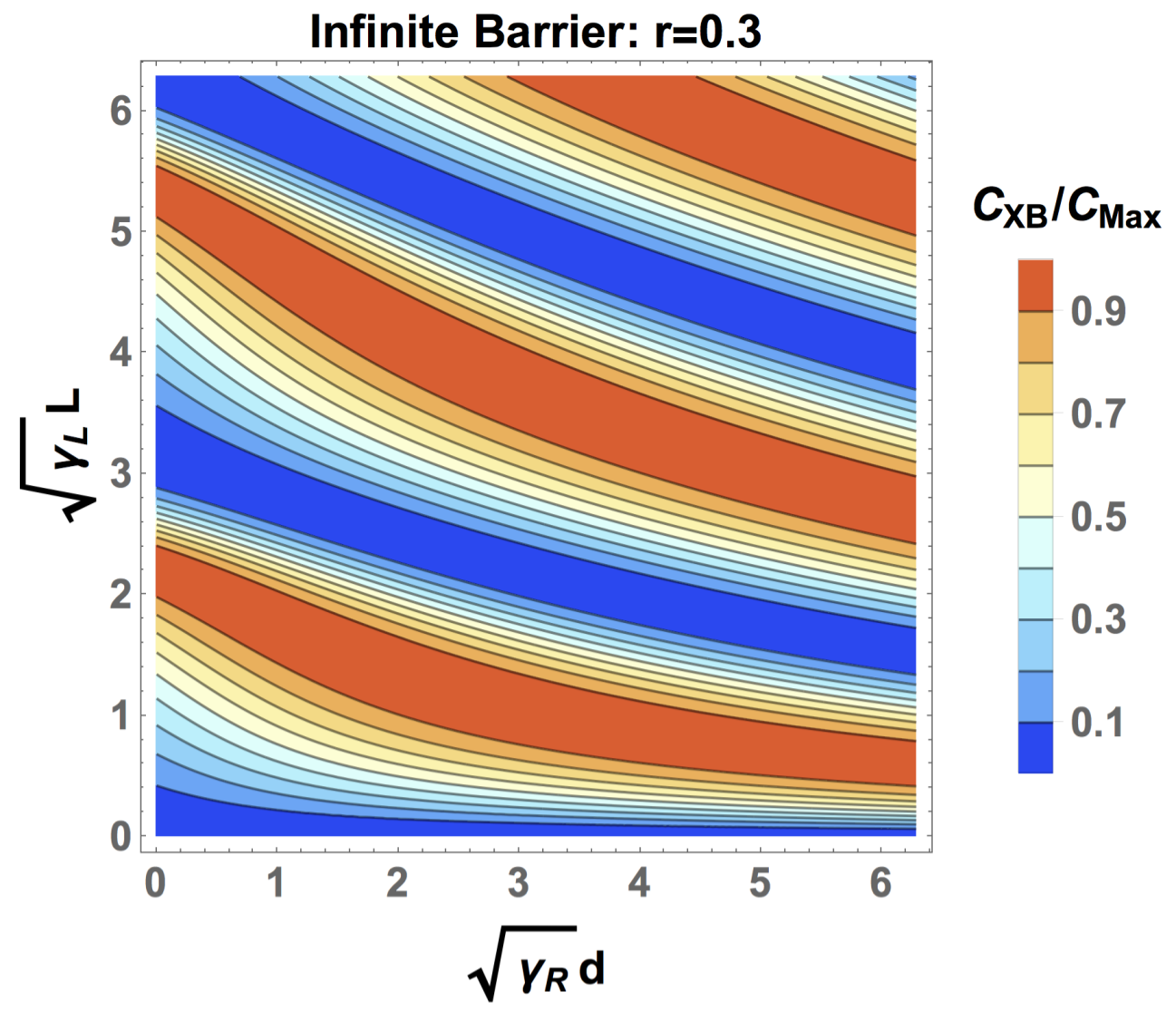}
\put(0,18){\large\textbf{(b)}}
\end{overpic}

\begin{overpic}[width=0.5\linewidth]{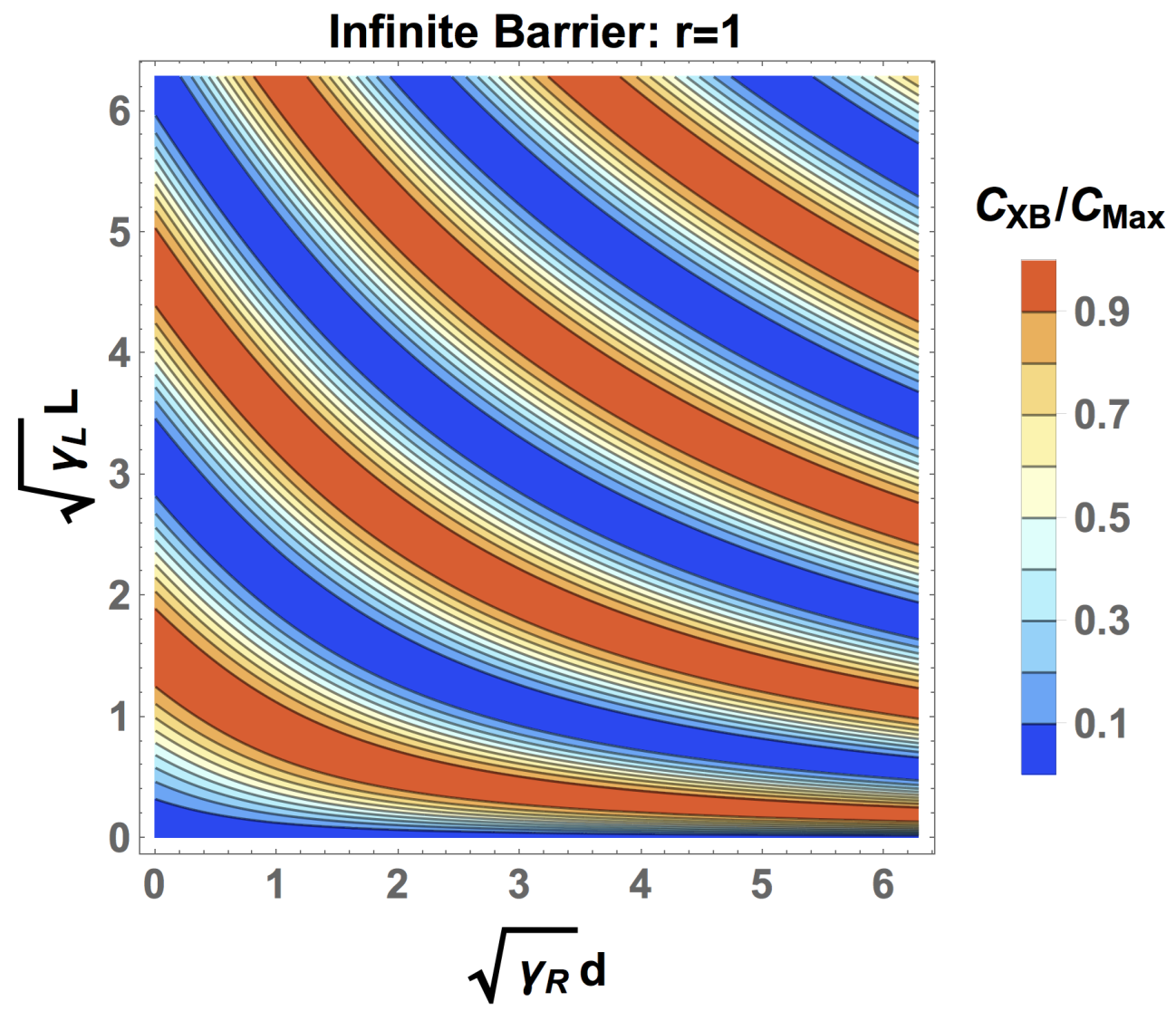}
\put(0,18){\large\textbf{(c)}}
\end{overpic}
\begin{overpic}[width=0.5\linewidth]{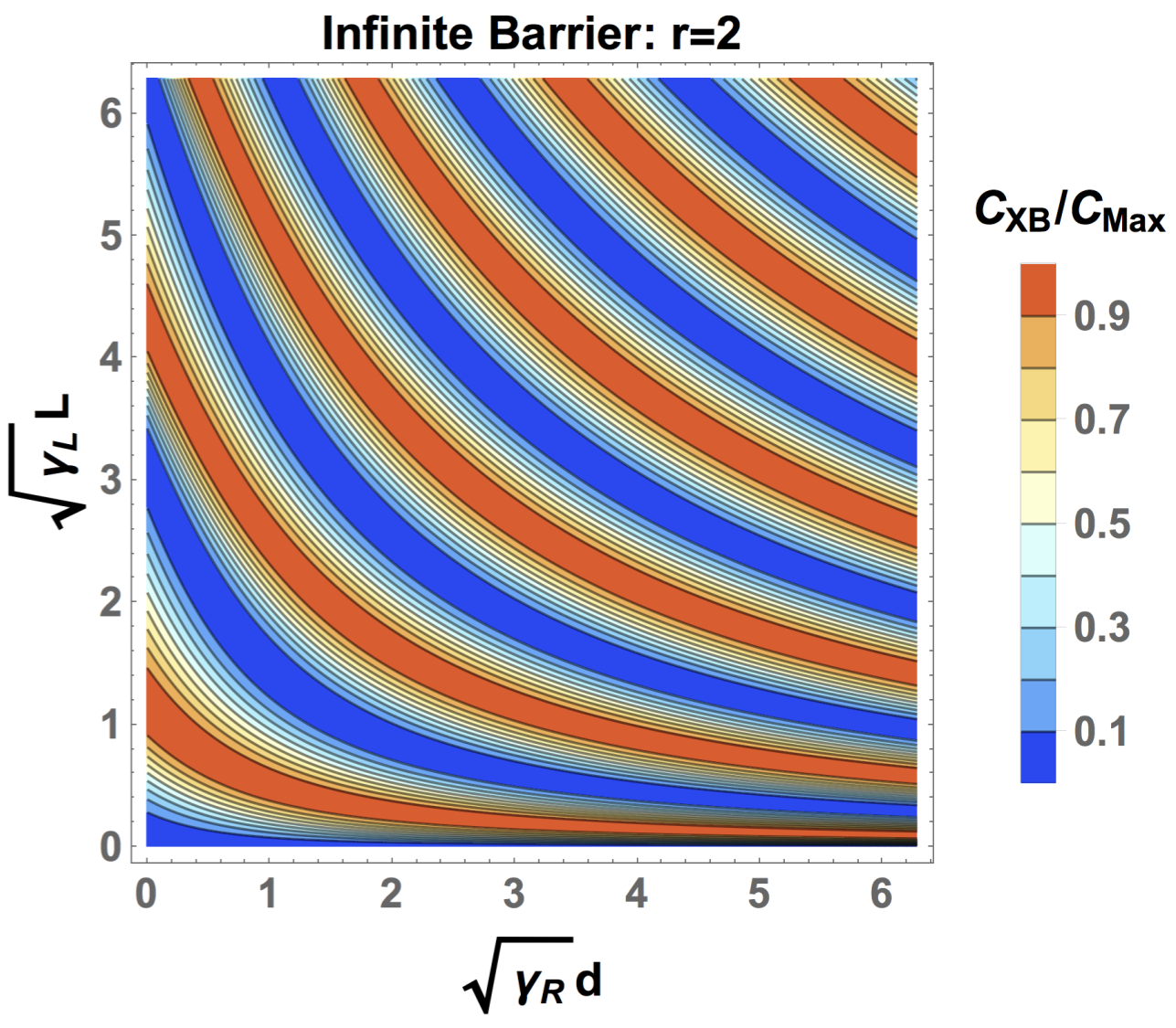}
\put(0,18){\large\textbf{(d)}}
\end{overpic}
\caption{Contour plots showing the variation of the oscillatory factor $\cos^2\left(\phi_\text{IB} \right)$ for the Infinite Barrier model of the X-B mode-conversion efficiency as the distance between the right-hand cutoff and the upper hybrid resonance ($d$), and the distance between the upper hybrid resonance and the left-hand cutoff ($L$) are varied. When $L$ and $d$ are specified, the Infinite Barrier model remains underdetermined via the additional asymmetry parameter $r$. In all plots, $r$ is constrained to take selected \emph{constant} values.}
\label{IB_CONT}
\end{figure}

Figure \ref{IB_CONT} is analogous to figure \ref{DC_CONT}; it shows the variation of the oscillatory factor $\cos^2\left(\phi_\text{IB} \right)$ for the IB model as the locations of the high-density and low-density cutoffs are varied. As with the DC model, the distance between the UHR and the RHC is normalized by the launched wavenumber as $\sqrt{\gamma_R}d$; unlike the DC model, however, the distance between the UHR and the LHC is normalized by the asymptotic \emph{wavenumber} of the HFS wave as $\sqrt{\gamma_L}L$, rather than the asymptotic decay coefficient. Also unlike the DC model, the IB model is not completely determined by fixing $\sqrt{\gamma_R}d$ and $\sqrt{\gamma_L}L$, as the asymmetry parameter $r$ remains unspecified. This figure features one possible convention of fixing $r$, namely that $r$ is a \emph{constant} with respect to the parameter scan; an alternate convention of fixing $r$ is discussed in the following section.

From figure \ref{IB_CONT}, it is clear that the IB phase model also exhibits a quasi-periodicity with respect to variation in $\sqrt{\gamma_L}L$ due to wave interference effects. However, when $r$ is constrained to be a constant of the parameter scan, the phase function can vary significantly with respect to $\sqrt{\gamma_R}d$. The extent of this variation is highly dependent on the fixed value of $r$: for small $r \ll 1$, the variation is minimal, while for $r \gtrsim 1$ the variation develops a hyperbolic character. Indeed, the density and curvature of the constructive interference `bands' in $\sqrt{\gamma_L}L$ - $\sqrt{\gamma_R}d$ space both increase as $r$ increases. Again, one observes that the IB model recovers the result $C_\text{XB}=0$ for $L=0$, while the result $C_\text{XB}=0$ for $d=0$ is guaranteed by the functional form of the envelope $C_\text{max}$.

Finally, as a concluding remark, it is constructive to consider $r$ as the additional free parameter in the IB model, as done here, rather than $L$ or $d$. This is because when fitting the IB model to experiment, $L$ and $d$ would be fixed by the experimentally-observed separation distances between the RHC, UHR, and LHC; $r$ would then serve as the sole fitting parameter. This will be discussed further in the following section.

\subsection{Model comparison}

At first glance, there is not much in common between the phase functions of the DC model and the IB model. Both models exhibit interference effects with respect to variation in $L$, and both models successfully recover the correct behavior for the confluent cases $L = 0$ and $d = 0$; however, the different characteristic frequencies of the quasi-periodic behavior of the two models implies that their difference can take any value on the interval $[0,1]$. This potentially large difference between the IB model and the DC model is fine if, for example, an experimentally-obtained X-B mode-conversion efficiency disagrees with that predicted by the DC model; in this case, it is a very good thing that an experimentalist have an alternative model to use. On the other hand, it is somewhat discomforting from a philosophical viewpoint that the two model descriptions of the same physical phenomenon can disagree so dramatically.

\begin{figure}
\begin{overpic}[width=0.5\linewidth]{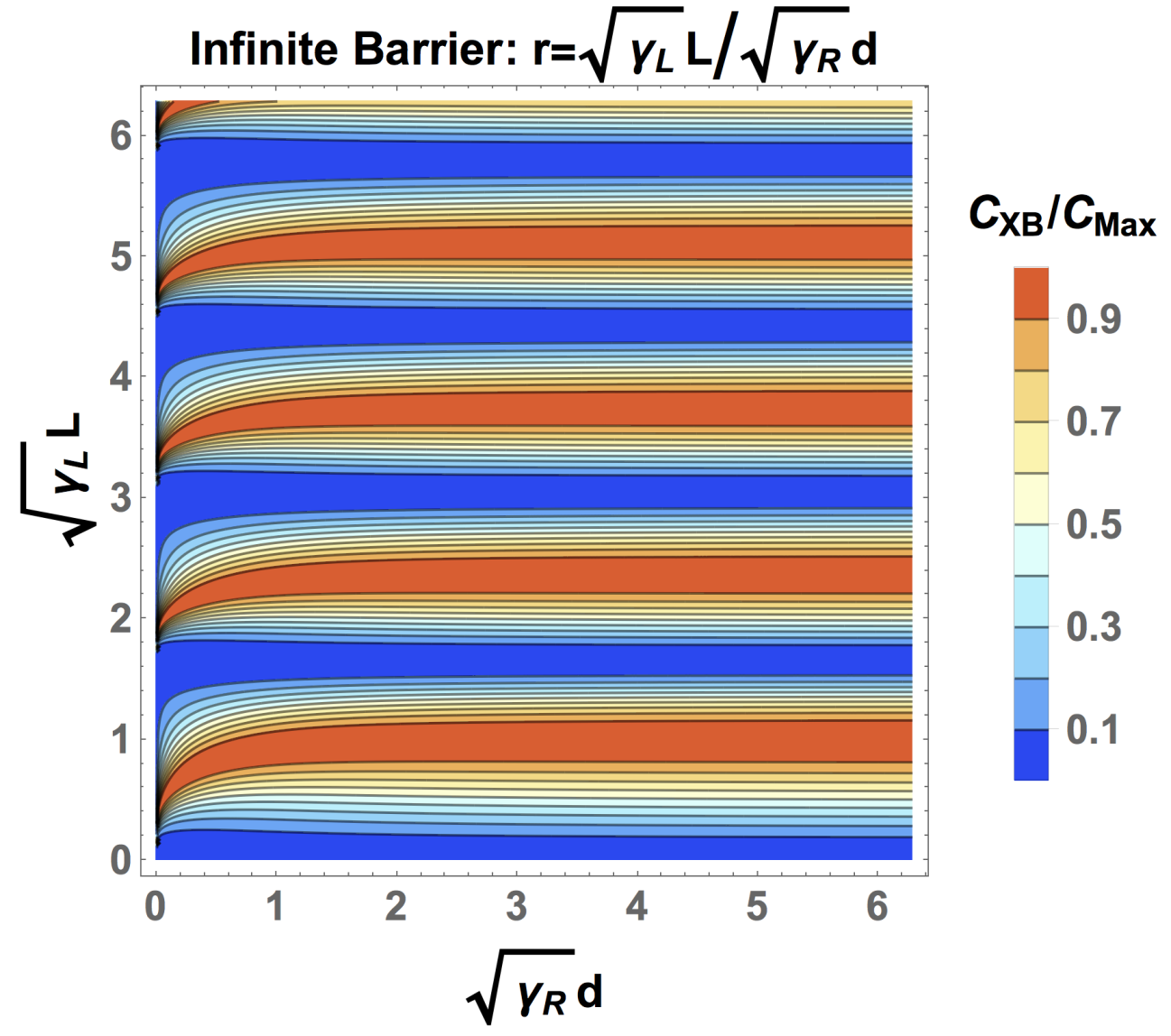}
\put(0,18){\large\textbf{(a)}}
\end{overpic}
\begin{overpic}[width=0.5\linewidth]{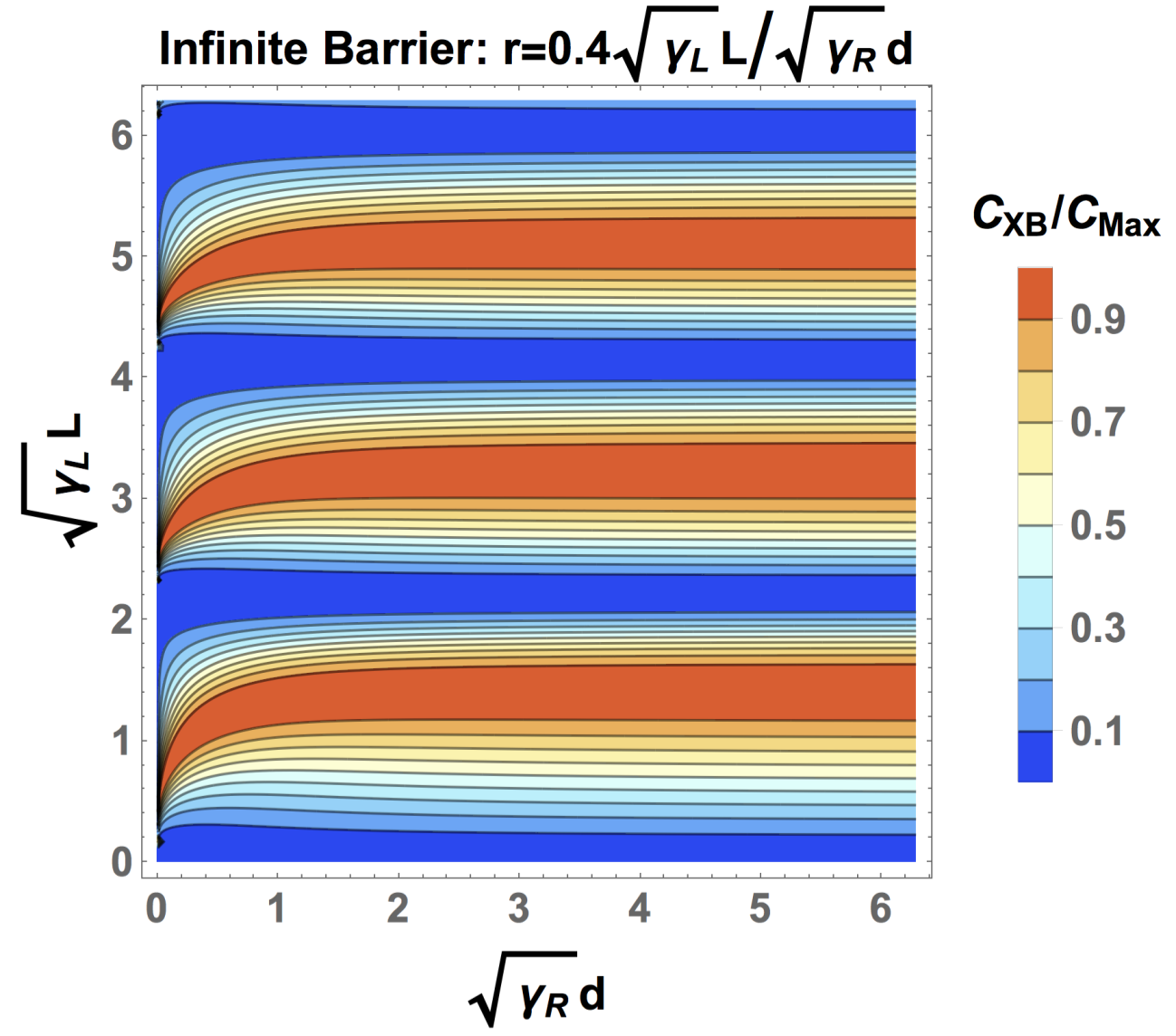}
\put(0,18){\large\textbf{(b)}}
\end{overpic}
\caption{Contour plots of the oscillatory factor $\cos^2\left(\phi_\text{IB} \right)$ of the Infinite Barrier model under the constraint $r=\alpha \hspace{0.5mm} r_\text{DC}$, where $r_\text{DC} \doteq \frac{\sqrt{\gamma_L}L}{\sqrt{\gamma_R}d}$. \textbf{(a)} When $\alpha = 1$, $r$ is constrained in an analogous manner to the DC model. \textbf{(b)} When $\alpha = 0.4$, the phase profiles of the two models become similar.}
\label{IB_constr}
\end{figure}

Fortunately, the two models can be brought into closer agreement by manipulating the convention of fixing $r$ in the IB model. An effective `$r$' for the DC model is defined by the fraction $\frac{\eta_L}{\eta_R} = \frac{\sqrt{\kappa_L}L}{\sqrt{\gamma_R}d}$. When the free parameters of the IB model are constrained analogously by the relation:
\begin{equation}
r_\text{DC}\doteq \frac{\sqrt{\gamma_L}L}{\sqrt{\gamma_R}d}
\label{rDC}
\end{equation}

\noindent the phase profiles of the two models become similar. Here, we have mapped $\kappa_L$ of the DC model to $\gamma_L$ of the IB model as the analogous parameter. The two phase profiles can be made nearly identical within some desired parameter range by manipulating this constraint on the IB model as $r=\alpha \hspace{0.5mm} r_\text{DC}$ for some $\alpha \in \mathbb{R}^+$. This possibility is explored in figure \ref{IB_constr}, which features the same plots as in figure \ref{IB_CONT} except with $r$ constrained to be a fixed \emph{function} of the parameter scan, namely $r = \alpha r_\text{DC}$. By comparing the two plots in figure \ref{IB_constr} with figure \ref{DC_CONT}, one sees that indeed the two models share the same qualitative behavior with this new constraint convention.

\begin{figure}
\begin{overpic}[width=0.5\linewidth]{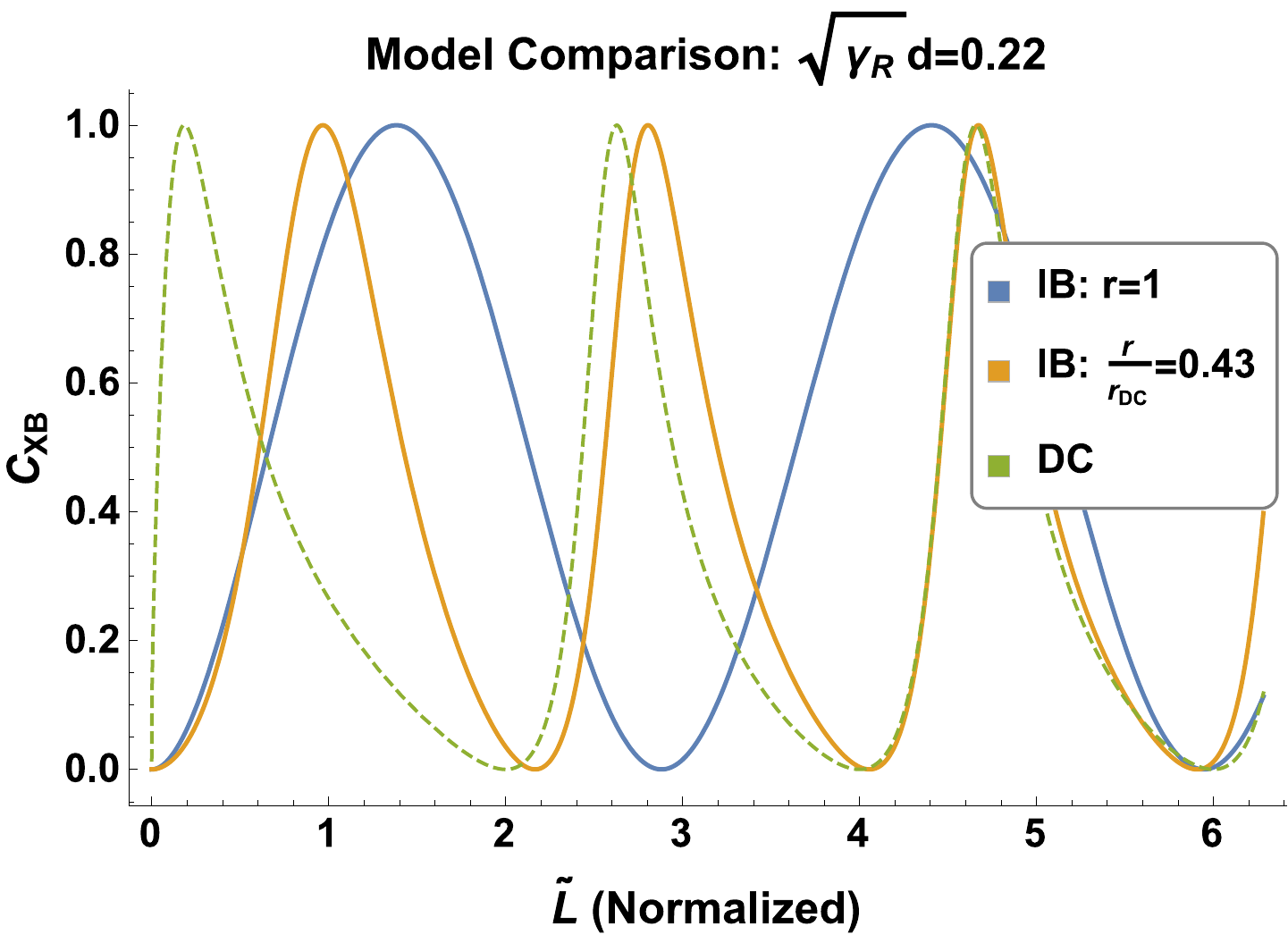}
\put(13,15){\large\textbf{(a)}}
\end{overpic}
\begin{overpic}[width=0.5\linewidth]{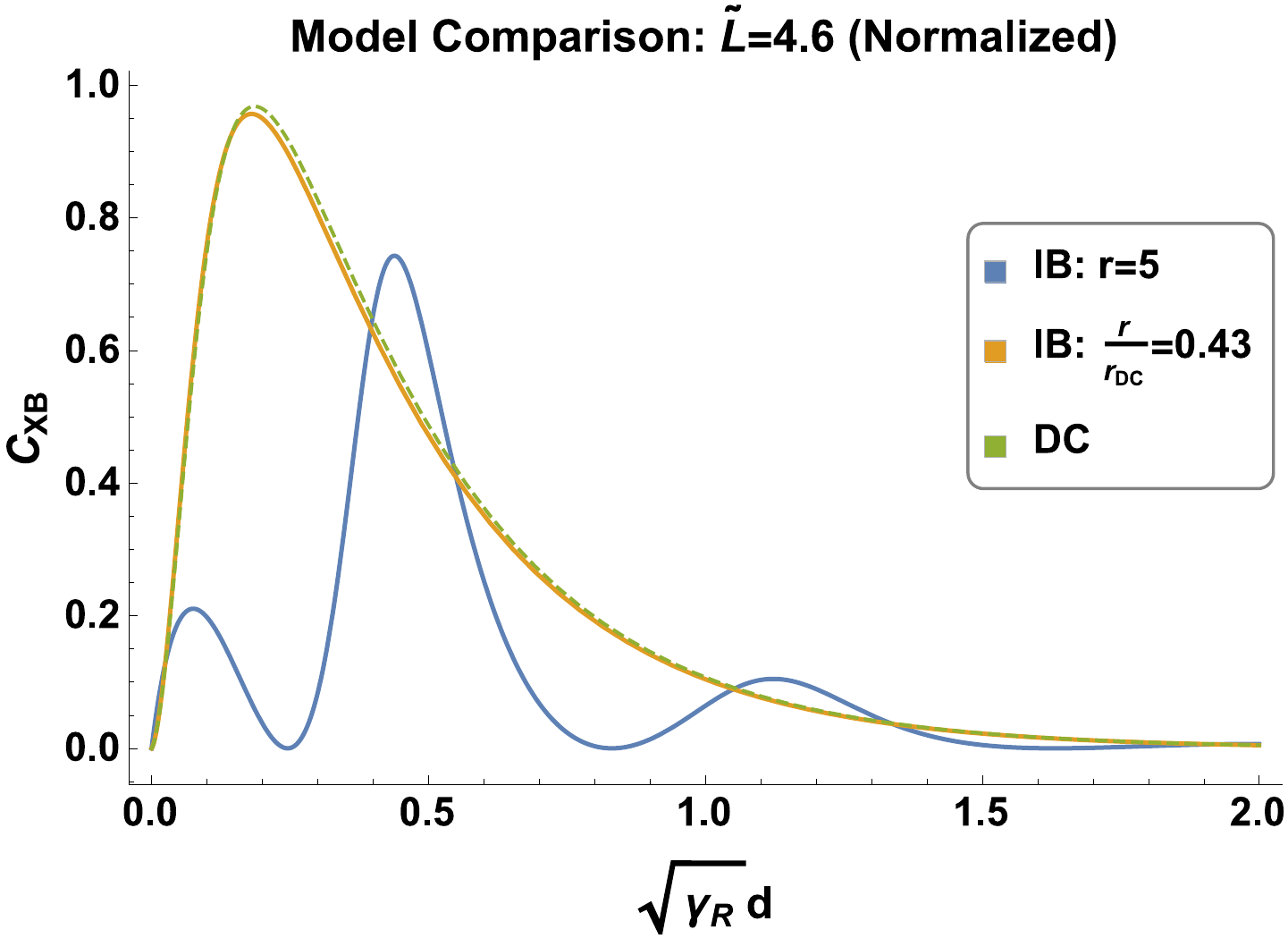}
\put(13,15){\large\textbf{(b)}}
\end{overpic}
\caption{Variation in the X-B mode-conversion efficiency for the Infinite Barrier and the Double-cutoff X-B phase models at \textbf{(a)} fixed $d$ and varying $L$, and at \textbf{(b)} fixed $L$ and varying $d$. In these plots, the free parameter $r$ of the IB model is constrained either to take a selected constant value, or to vary analogously with the DC model constraint. We use the notation $\tilde{L}$ to denote generically that $L$ is normalized according to the prescriptions used in figures \ref{DC_CONT} and \ref{IB_CONT}, that is, $\tilde{L} = \sqrt{\kappa_L}L$ for the DC model, and $\tilde{L} = \sqrt{\gamma_L}L$ for the IB model.}
\label{modelCOMP}
\end{figure}

Figure \ref{modelCOMP} further compares the IB model and the DC model at select parameters, and for both conventions of fixing the additional parameter $r$ of the IB model. For convenience, we denote the normalized distance between the LHC and the UHR generically as $\tilde{L}$; it is understood that the normalization is performed in accordance with the conventions established in figures \ref{DC_CONT} and \ref{IB_CONT}, namely $\tilde{L} = \sqrt{\kappa_L}L$ for the DC model, and $\tilde{L} = \sqrt{\gamma_L}L$ for the IB model. The first plot of this figure compares the IB model with the DC model when only $\tilde{L}$ is allowed to vary. This plot elucidates the statement made in the previous paragraph: selecting $\alpha = 0.43$ in the constraint $r = \alpha r_\text{DC}$ brings the IB model and the DC model into close agreement for the parameter range $\sqrt{\gamma_R}d = 0.22$, $\tilde{L} \in [4,6]$. However, with $\alpha = 0.43$, the two models do not agree on the interval $\tilde{L} \in (0,2]$, for example. To bring the two models into agreement within this parameter regime would require a different choice of $\alpha$.

On the other hand, it may be advantageous to fully utilize the additional free parameter of the IB model. Naturally, this additional degree of freedom allows the IB model to exhibit behavior that is impossible within the DC model. For example, as shown in the second plot of figure \ref{modelCOMP}, the IB model permits multi-peaked behavior of the mode-conversion efficiency with respect to variations in $\sqrt{\gamma_R}d$ at fixed $\sqrt{\gamma_L}L$ when $r$ is a constant. When $r$ is constrained analogously to the DC model, the IB model and the DC model agree to remarkable accuracy. Thus, the IB model can exhibit features beyond the capacity of the DC model while reproducing the DC model behavior when appropriately constrained.

An important role for any theoretical model is to provide experimental predictions based on input plasma parameters. For our case here, the necessary experimental parameters are: (1) the launched wave frequency $\omega$, (2) the experimentally-observed distance between the RHC and the UHR $d_\text{exp}$, and (3) the experimentally-observed distance between the UHR and the LHC $L_\text{exp}$. Recall that $\omega$ is needed because both models use normalized coordinates. The DC model is completely specified by these 3 experimental parameters, by setting (1) $\gamma_R = 1$ for vacuum-launch, (2) $\beta = \gamma_R \cdot d_\text{exp}\frac{\omega}{c}$, and (3) $\gamma_L = \frac{\beta}{L_\text{exp}}\cdot \frac{c}{\omega}$. The model parameters $\eta_R$ and $\eta_L$ are computed from these three quantities by their definitions. In contrast, the IB model is not fully specified by these 3 experimental parameters. Indeed, one sets (1) $\gamma_R = 1$ for vacuum-launch, (2) $\beta = \gamma_R \cdot d_\text{exp}\frac{\omega}{c}$, and (3) $L = L_\text{exp}\frac{\omega}{c}$. However, $\gamma_L$ remains a free parameter that can be used to fit the IB model potential more accurately to the experimentally-observed index of refraction profile than the DC model potential.

Alternatively, the remaining free parameter $\gamma_L$ can be set by providing an additional experimental measurement, such as an experimentally measured X-mode reflection coefficient. Suppose an X-B experiment is performed with parameters $\omega$, $d_\text{exp}$, and $L_\text{exp}$, and a reflection coefficient of $R_\text{exp}$ is measured. Then, $\gamma_L$ is computed as $C_\text{IB}\left(\beta = d_\text{exp}\frac{\omega}{c}, \gamma_R=1, \gamma_L, L=L_\text{exp}\frac{\omega}{c} \right) = 1-R_\text{exp}$ using a simple 1-D root-finding method, where $C_\text{IB}$ is computed using equations \ref{convEFF} and \ref{IBphase}. With the IB model fully specified, an experimentalist can then determine how the plasma equilibrium should be tuned to improve the X-B mode-conversion efficiency.

Often, however, the locations of the cutoffs and resonances are not known precisely, due to the presence of blobs and so forth. In this case, it can be beneficial that the IB model, by construction, does not resolve the field behavior near the high-density cutoff. One can choose an approximate value for $L_\text{exp}$, and then use the additional free parameter $\gamma_L$ to slightly correct any loss in accuracy via the additional fitting step. As will be shown in the following section, this type of procedure can be useful for assessing fluctuating plasmas, in which $L_\text{exp}$ need only be an estimated mean value.

For these reasons, the IB model is the advantageous choice of the pair when fitting to experimental data. The paradigm that the X-B problem can be formulated as a boundary-value problem on a finite domain is also readily generalizable to higher-dimensional computational studies. A computer code based on the IB model should be able to quickly diagnose an X-B experiment, and  inform the experimenters on any necessary adjustments to improve the mode-conversion efficiency. Such a tool will be particularly useful in experiments where the 1-D analytical expressions for the X-B mode-conversion efficiency are not suitable, such as with the strongly-shaped and strongly-sheared equilibrium of an ST.

As a final remark, in the previous discussion we consider using the plasma equilibrium parameters to predict the X-B mode-conversion efficiency. In principle, one can also use a subset of the plasma equilibrium parameters, along with a measured reflection coefficient, to infer the remaining plasma equilibrium parameters. For example, provided $\omega$, $d_\text{exp}$, and $R_\text{exp}$, one can compute $L_\text{exp}$ using a selected model of the X-B mode-conversion efficiency. In this form, an X-B experiment becomes akin to a reflectometry experiment, although the high sensitivity of the X-B mode-conversion efficiency to model details means that such an application should be used with caution.

\section{Effect of small-amplitude density fluctuations}
\label{fluctuation}

The X-B mode-conversion models presented in the previous section assume a very specific form for the density and magnetic field profiles to simplify the analysis. In a more realistic model, the density and magnetic fields will be contaminated with micro-turbulent fluctuations. These fluctuations are typically very slow (kHz - MHz range) compared to EC wave timescales (GHz range), so they affect the X-mode propagation primarily as a time-independent modification to the background plasma. Moreover, in most tokamak experiments, the magnetic field fluctuations are negligibly small, with amplitudes on the order of $\frac{\delta B}{B} \sim 10^{-5}$. Density fluctuation amplitudes, in contrast, are on the order of $\frac{\delta n}{n} \sim 0.1$, sometimes approaching or exceeding unity when blobs are present\cite{Zweben07}. We therefore consider only small-amplitude, stationary density fluctuations in the forthcoming analysis.

When the density fluctuations are small, their effects on the X-B mode-conversion efficiency can be obtained analytically in 1-D using a perturbative approach and a Green's function\cite{Morse53}. Such analysis is possible because both the DC and the IB models formulate the X-B mode-conversion as the boundary-value solution to a specific differential equation; the solution to either model is therefore readily incorporated into a Green's function that can be used to study more complicated equilibria. Being defined on a finite domain, the IB model is considerably simpler to analyze computationally than the DC model. Since a range of validity can be placed on the perturbative results only after a comparison with numerical simulations, only the IB model will be used in this section.

Let the density profile be given as $n(x) = n_0(x) + \delta n(x)$, where $\delta n(x)$ is the perturbation to the equilibrium state defined by $n_0(x)$. By assumption, these density perturbations are small, such that there exists a small parameter $\epsilon$ defined as:
\begin{equation}
\epsilon = \max_{x\in \mathbb{R}}~\frac{\delta n(x)}{n_0(x)} \ll 1
\end{equation}

\noindent To this end, let $\tilde{n}(x)$ be an $O(1)$ function such that:
\begin{equation}
n(x) = n_0(x)\left(1 + \epsilon \tilde{n}(x) \right)
\end{equation}

\noindent Since $S(x) - 1$ and $D(x)$ are both homogeneous functions of degree 1 in $n(x)$, $S(x)$ and $D(x)$ can be expressed in terms of $\epsilon$ as:
\numparts
\begin{eqnarray}
S(x) = S_0(x) + \epsilon(S_0(x) - 1)\tilde{n}(x)\\
D(x) = D_0(x) + \epsilon D_0(x) \tilde{n}(x)
\end{eqnarray}
\endnumparts

\noindent where $S_0(x)$ and $D_0(x)$ are the limiting forms of $S(x)$ and $D(x)$ when there are no density perturbations. Note that the above expressions for $S(x)$ and $D(x)$ are \emph{exact}; there are no approximations performed yet.

The $\epsilon = 0$ limit is chosen to coincide with the Budden equation. To facilitate this, we first define:
\begin{equation}
\gamma(x) \doteq \gamma_R\left[\frac{1}{r^2} +\Theta(x)\cdot\left(1-\frac{1}{r^2} \right)\right] = \left\{\begin{array}{c c}
\gamma_R \cdot r^{-2} & x < 0\\
\frac{1}{2}\gamma_R\cdot\left(1+r^{-2}\right) & x = 0\\
\gamma_R & x > 0
\end{array} \right.
\end{equation}

\noindent with $\gamma_R$ and $r$ both real and positive, and $\Theta(x)$ is the Heaviside step function. Then, the Budden potential is obtained in the $\epsilon = 0$ limit of equation \ref{potential} when $S_0(x)$ and $D_0(x)$ have the following forms:
\numparts
\begin{eqnarray}
S_0(x) = \frac{\gamma^2(x)}{4\beta}x\\
D_0(x) = \frac{\gamma^2(x)}{4\beta}\left(x - \frac{2\beta}{\gamma(x)} \right)
\end{eqnarray}
\endnumparts

\noindent With the functions $S(x)$ and $D(x)$ now fully specified, the X-mode wave equation in the presence of density fluctuations is given as:
\numparts
\begin{eqnarray}
\label{fluctBUDDENeq}
E_y''(x) = -\tilde{Q}(x)E_y(x)\\
\tilde{Q}(x) = \frac{\gamma(x)x-\beta + 2\epsilon \tilde{n}(x)\left(\frac{}{}[\gamma(x) - 1]x - \beta\right)}{x+\epsilon\tilde{n}(x)\left(x - \frac{4\beta}{\gamma^2(x)} \right)}
\label{fluctBUDDENQ}
\end{eqnarray}
\endnumparts

\begin{figure}
\begin{overpic}[width=0.5\linewidth]{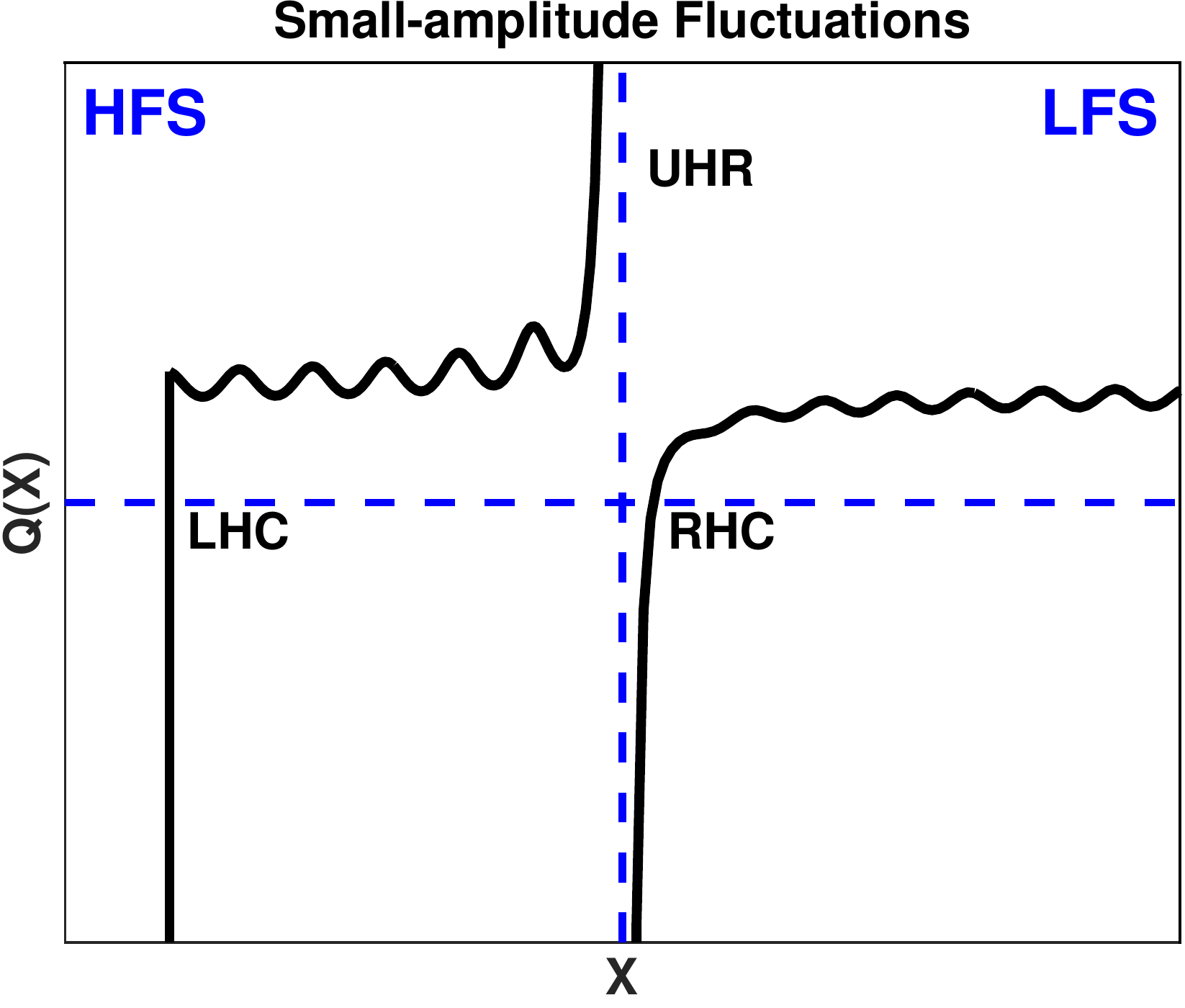}
\put(5,10){\large\textbf{(a)}}
\end{overpic}
\begin{overpic}[width=0.5\linewidth]{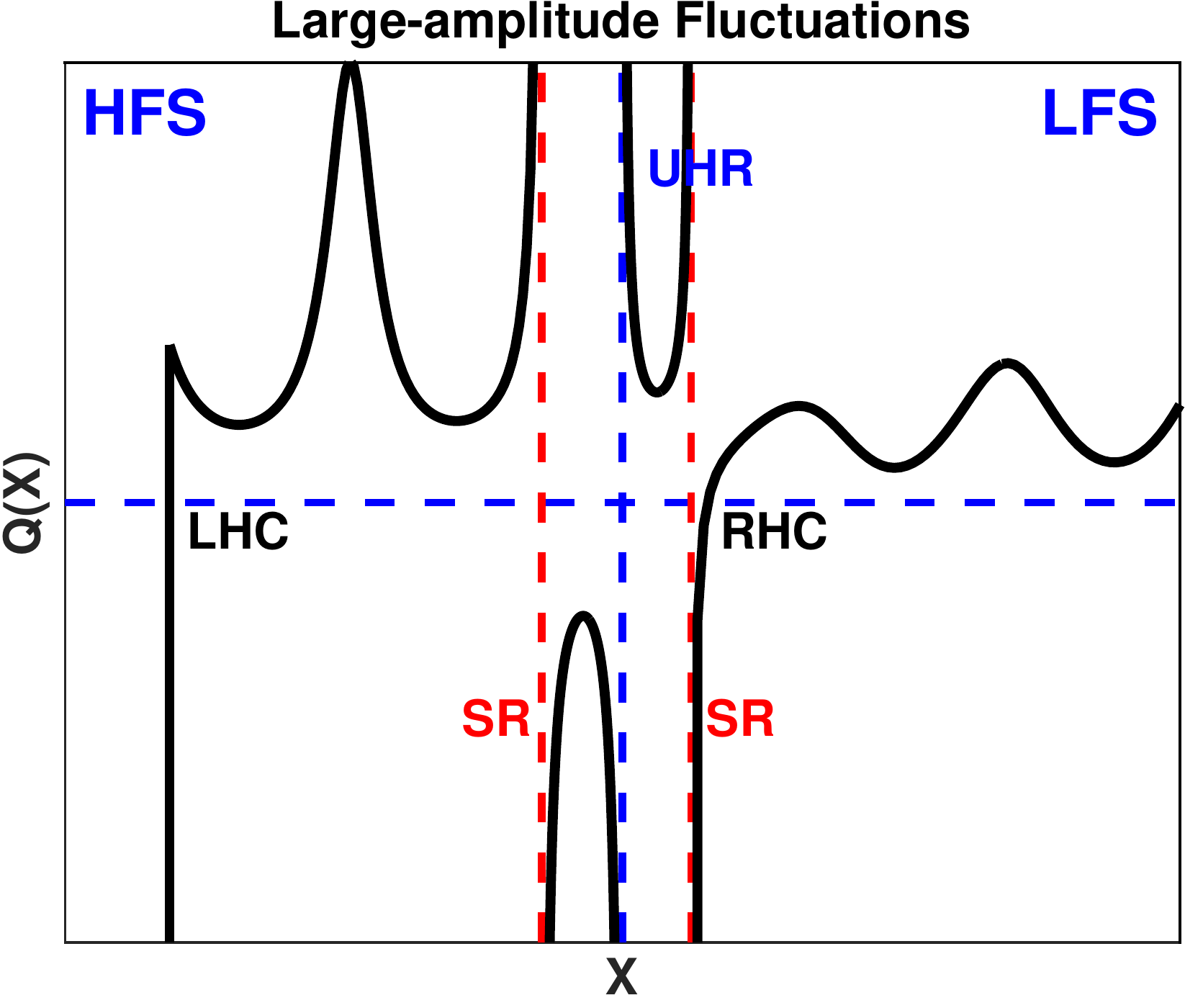}
\put(5,10){\large\textbf{(b)}}
\end{overpic}
\caption{The Infinite Barrier model potential function in the presence of sinusoidal density fluctuations. \textbf{(a)} Small-amplitude density fluctuations provide a small, periodic modification to the potential function. Here, $\epsilon = 0.1$. \textbf{(b)} Large-amplitude density fluctuations can yield substantial alterations to the potential function topology, introducing additional side-resonances (SRs) and cutoffs. Here, $\epsilon = 0.5$.}
\label{FluctQ}
\end{figure}

\noindent Again, this equation is \emph{exact}.

Figure \ref{FluctQ} shows the modified IB model potential function in the presence of small-amplitude and large-amplitude density fluctuations. The modified IB model potential function corresponds to $\tilde{Q}(x)$ on the domain $x \in [-L,\infty)$, and $-\infty$ for $x < -L$. As seen from the figure, the amplitude of the density perturbations can have profound consequences on the topology of the modified potential function. When the fluctuations are `small', the modification to the potential function is also small, and, for the most part, can be treated with perturbation techniques. 

In contrast, when the fluctuation amplitudes are `large', the potential function is modified substantially. The fluctuations can create additional cutoffs and resonances, which we term side-resonances (SRs). These significantly modify the cutoff-resonance-cutoff triplet structure of the IB model, and correspondingly, modify the X-B mode-conversion physics. Although we mention this large-amplitude regime here for completeness, this regime is difficult to study analytically and numerically, and will be the focus of future work. Among other reasons, the presence of multiple resonances means that the X-mode amplitude sink is no longer uniquely specified - it is no longer obvious how the mode-converted X-mode amplitude is partitioned among the different EBW branches. The `resonant absorption' technique for computing the mode-conversion efficiency is no longer valid, and only a fully-kinetic treatment can resolve this conundrum.

Let us now confine ourselves to the `small-amplitude' regime and adopt a perturbative approach. Let us assume the solution to equation \ref{fluctBUDDENeq} can be expressed as a power series in $\epsilon$ as:
\begin{equation}
E_y(x) \approx y_0(x) + \epsilon y_1(x) + O(\epsilon^2)
\end{equation}

\noindent For ease of notation, we have introduced the functions $y_j$ to denote the $O(\epsilon^j)$ term in the expansion of $E_y(x)$. Then, $y_0(x)$ satisfies the Budden equation with the IB boundary condition, whose solutions have been discussed at length in the previous section. 

Let us consider the $O(\epsilon)$ equation which governs $y_1(x)$:
\numparts
\begin{eqnarray}
\label{y1ODE}
y_1'' + \left(\gamma(x) - \frac{\beta}{x}\right) y_1 = f(x)\\
f(x) = \left[\frac{4\beta^2}{\gamma^2(x)} +x\beta\left(1-\frac{4}{\gamma(x)} \right)+x^2 \left(\frac{}{}2-\gamma(x) \right)\right]\frac{\tilde{n}(x)}{x^2} y_0(x)
\label{fRHS}
\end{eqnarray}
\endnumparts

\noindent In writing equations \ref{y1ODE} and \ref{fRHS}, we have used the fact the $y_0(x)$ is a solution to the IB model differential equation:
\begin{equation}
y_0(x) = B\left\{\begin{array}{l l}
\tilde{A} W_{-i\frac{\eta}{2}}\left(2\sqrt{\gamma_R}x e^{-i\frac{\pi}{2}}\right) + W_{i\frac{\eta}{2}}\left(2\sqrt{\gamma_R}x e^{-i\frac{3\pi}{2}}\right) & x > 0\\
\tilde{C} W_{-i\frac{\eta r}{2}}\left(2\frac{\sqrt{\gamma_R}}{r}x e^{-i\frac{\pi}{2}}\right) + \tilde{D} W_{i\frac{\eta r}{2}}\left(2\frac{\sqrt{\gamma_R}}{r}x e^{-i\frac{3\pi}{2}}\right) & x \le 0
\end{array}\right.
\end{equation}

\noindent where $\tilde{A}$, $\tilde{C}$, and $\tilde{D}$ are the solutions to equation \ref{linSys}. Clearly, $y_1(x)$ can be expressed in terms of the Green's function to the Budden equation, which is derived in the following subsection.

\subsection{The Green's function to the Budden equation}

Let $G(x,\xi)$ be the Green's function for the Budden equation. Then, $G(x,\xi)$ satisfies:
\begin{equation}
G''(x,\xi) + \left(\gamma(x) - \text{PV}\left[\frac{\beta}{x}\right]\right)G(x,\xi) = \delta(x-\xi) - i\pi \beta \delta(x) G(x,\xi)
\label{GODE}
\end{equation}

\noindent where prime denotes $\frac{\partial}{\partial x}$, $\text{PV}$ denotes the principal value, $\delta(x)$ denotes the Dirac delta function, and we have used the Sokhotskii-Plemelj representation for $\frac{1}{x}$ assuming a vanishingly small negative imaginary component to the pole\cite{Sokhotskii73,Plemelj08}:
\begin{equation}
\lim_{\epsilon\to 0^+} \frac{1}{x+i\epsilon} = \text{PV}\left[\frac{1}{x} \right] - i\pi \delta(x)
\end{equation}

The Sokhotskii-Plemelj representation is equivalent to the choice of branch cut used in the previous section; however, using this representation makes it straightforward to derive the generalized continuity conditions for a Green's function to a singular differential equation. The continuity conditions for the Green's function solution to equation \ref{GODE} are:
\numparts
\begin{eqnarray}
\label{GcontS}
G(0^+,\xi) = G(0^-,\xi)\\
G(\xi^+,\xi) = G(\xi^-,\xi)\\
\hspace{-2mm}\left.\begin{array}{l}
G'(0^+,\xi) = G'(0^-,\xi) - i\pi\beta G(0,\xi)\\
G'(\xi^+,\xi) = G'(\xi^-,\xi) + 1
\end{array}\right\}: \xi \neq 0\\
\hspace{-2mm}\left.~G'(0^+,0) = G'(0^-,0) + 1 - i\pi\beta G(0,0)\right\}: \xi = 0
\label{GcontE}
\end{eqnarray}
\endnumparts

\noindent These continuity conditions must be supplemented with model-specific boundary conditions. For the IB model, the relevant boundary condition is:
\begin{equation}
G(-L,\xi) = 0
\end{equation}

\noindent which places an infinite barrier at $x=-L$. 

Let us first consider the case $\xi \neq 0$. Then, since $G(x,\xi)$ is a solution to the Budden equation for every $x\neq \xi$, $G(x,\xi)$ can be expressed in terms of Whittaker functions as:
\numparts
\begin{eqnarray}
\hspace{-19mm}
G(x,\xi > 0) = \left\{\begin{array}{l l}
A_+ W_{-i\frac{\eta}{2}}\left(2\sqrt{\gamma_R}xe^{-i\frac{\pi}{2}}\right) + B_+ W_{i\frac{\eta}{2}}\left(2\sqrt{\gamma_R}xe^{-i\frac{3\pi}{2}}\right) & x > \xi\\
C_+ W_{-i\frac{\eta}{2}}\left(2\sqrt{\gamma_R}xe^{-i\frac{\pi}{2}}\right) + D_+ W_{i\frac{\eta}{2}}\left(2\sqrt{\gamma_R}xe^{-i\frac{3\pi}{2}}\right) & 0 < x < \xi\\
F_+ W_{-i\frac{\eta r}{2}}\left(2\frac{\sqrt{\gamma_R}}{r}xe^{-i\frac{\pi}{2}}\right) + H_+ W_{i\frac{\eta r}{2}}\left(2\frac{\sqrt{\gamma_R}}{r}xe^{-i\frac{3\pi}{2}}\right) & x < 0
\end{array}\right.\\
\hspace{-19mm}
G(x,\xi < 0) = \left\{\begin{array}{l l}
A_- W_{-i\frac{\eta}{2}}\left(2\sqrt{\gamma_R}xe^{-i\frac{\pi}{2}}\right) + B_- W_{i\frac{\eta}{2}}\left(2\sqrt{\gamma_R}xe^{-i\frac{3\pi}{2}}\right) & x > 0\\
C_- W_{-i\frac{\eta r}{2}}\left(2\frac{\sqrt{\gamma_R}}{r}xe^{-i\frac{\pi}{2}}\right) + D_- W_{i\frac{\eta r}{2}}\left(2\frac{\sqrt{\gamma_R}}{r}xe^{-i\frac{3\pi}{2}}\right) & \xi < x < 0\\
F_- W_{-i\frac{\eta r}{2}}\left(2\frac{\sqrt{\gamma_R}}{r}xe^{-i\frac{\pi}{2}}\right) + H_- W_{i\frac{\eta r}{2}}\left(2\frac{\sqrt{\gamma_R}}{r}xe^{-i\frac{3\pi}{2}}\right) & x < \xi
\end{array}\right.
\end{eqnarray}
\endnumparts

\noindent As in the previous section, the continuity and boundary requirements of $G(x,\xi)$ can be written as linear systems for the coefficients. For $\xi > 0$, the linear system of interest is:
\begin{equation}
\tilde{\textbf{A}}_+
=\textbf{b}_+ \textbf{M}_{(+)}^{-1}
\label{linPLUS}
\end{equation}

\noindent while for $\xi < 0$, the linear system of interest is:
\begin{equation}
\tilde{\textbf{A}}_-
=\textbf{b}_-\textbf{M}_{(-)}^{-1}
\label{linMIN}
\end{equation}

\noindent where we have defined the vector of coefficients:
\numparts
\begin{eqnarray}
\tilde{\textbf{A}}_+ \doteq \left(\begin{array}{c c c c c}
\tilde{A}_+, & \tilde{C}_+, & \tilde{D}_+, & \tilde{F}_+, & \tilde{H}_+
\end{array}\right)\\
\tilde{\textbf{A}}_- \doteq \left(\begin{array}{c c c c c}
\tilde{A}_-, & \tilde{C}_-, & \tilde{D}_-, & \tilde{F}_-, & \tilde{H}_-
\end{array}\right)
\end{eqnarray}
\endnumparts

\noindent As before, the tilde denotes normalization with respect to the incident X-mode wave amplitude: $\tilde{X}_+ \doteq \frac{X_+}{B_+}$, and $\tilde{X}_- \doteq \frac{X_-}{B_-}$ for $X \in \{A,C,D,F,H \}$. The matrices $\textbf{M}_{(\pm)}$ and the vectors $\textbf{b}_{\pm}$ are shown in \ref{IBGreen}.

Next, let us consider the case $\xi = 0$. Now, $G(x,0)$ can be expressed as:
\begin{equation}
G(x,0) = \left\{\begin{array}{l l}
A_0 W_{-i\frac{\eta}{2}}\left(2\sqrt{\gamma_R}x e^{-i\frac{\pi}{2}}\right) + B_0 W_{i\frac{\eta}{2}}\left(2\sqrt{\gamma_R}x e^{-i\frac{3\pi}{2}}\right) & x > 0\\
C_0 W_{-i\frac{\eta r}{2}}\left(2\frac{\sqrt{\gamma_R}}{r}x e^{-i\frac{\pi}{2}}\right) + D_0 W_{i\frac{\eta r}{2}}\left(2\frac{\sqrt{\gamma_R}}{r}x e^{-i\frac{3\pi}{2}}\right) & x < 0
\end{array}\right.
\end{equation}

\noindent The continuity and boundary conditions then lead to the linear system:
\begin{equation}
\tilde{\textbf{A}}_0
=\textbf{b}_0\textbf{M}_{(0)}^{-1}
\label{linZERO}
\end{equation}

\noindent where the matrix $\textbf{M}_{(0)}$ and the vector $\textbf{b}_0$ are presented in \ref{IBGreen}, while the vector of coefficients is defined as:
\begin{equation}
\tilde{\textbf{A}}_0 \doteq \left(\begin{array}{c c c}
\tilde{A}_0, & \tilde{C}_0, & \tilde{D}_0
\end{array}\right)
\end{equation}

\noindent Again, $\tilde{X}_0 \doteq \frac{X_0}{B_0}$ for $X\in\{A, C, D \}$.

\subsection{The modified X-B reflection coefficient}

For algebraic convenience, let us specialize to consider only the effect of quasi-monochromatic density fluctuations on the X-B mode-conversion efficiency. Such density fluctuations are described by the family of functions given as:
\begin{equation}
\tilde{n}(x;k_n,L_n) = \frac{\sqrt{2e}}{L_n}x\sin(k_nx)e^{-\frac{x^2}{L_n^2}}
\label{densFLUCT}
\end{equation}

\noindent where $k_n$ and $L_n$ denote the characteristic wavelength and the characteristic decay length of the density fluctuations. Physically, choosing the density perturbation to vanish at the origin means that there is no shift to the location of the UHR. This procedure can always be performed without loss of generality by properly partitioning the `equilibrium' and `fluctuation' density profiles. In other words, all density perturbations are defined with respect to the density at the UHR. The Gaussian envelope is included in equation \ref{densFLUCT} to assess only the impact of the density fluctuations in the immediate vicinity of the mode-conversion region. The factor of $x$ multiplying the sinusoidally-modulated Gaussian is introduced to ensure that the driving term $f(x)$ in equation \ref{fRHS} is analytic everywhere. This is a stronger constraint on $\tilde{n}(x)$ than simply requiring $\tilde{n}(x)$ be analytic. Finally, the overall constant factor $\frac{\sqrt{2e}}{L_n}$ is the normalization to ensure that the envelope has a maximum value of $1$.

The general case for arbitrary analytic $\tilde{n}(\xi)$ that vanishes at the origin is discussed in \ref{modREF}. Here, we shall simply make use of the results therein discussed, noting that integrals of the form $\int_x^\infty d\xi$ are subdominant to those of the form $\int_0^\infty d\xi$ due to the exponential decay of the integrand. Hence, asymptotically as $x\to \infty$, $y_1(x)$ is given by the expression:
\begin{eqnarray}
y_1(x) \sim y_{1,\text{in}} y_-(x) + y_{1,\text{out}}y_+(x)
\end{eqnarray}
\numparts
\begin{eqnarray}
y_{1,\text{in}} = -\mathcal{F}^-_{-L}(0)\\
y_{1,\text{out}} = \frac{2\pi ie^{-\pi\frac{\eta}{2}}}{\Gamma\left(-i\frac{\eta}{2} \right)\Gamma\left(1-i\frac{\eta}{2} \right)}\mathcal{F}^-_{-L}(0)-e^{-\pi\eta}\mathcal{A}^-_{-L}(0)
\end{eqnarray}
\endnumparts

\noindent where we have defined the expressions $\mathcal{F}^-_{-L}(0) \doteq -\int_{-L}^0 f(\xi)d\xi - \int_0^\infty f(\xi) d\xi$, and $\mathcal{A}^-_{-L}(0) \doteq -\int_{-L}^0 \tilde{A}_-(\xi)f(\xi)d\xi - \int_0^\infty \tilde{A}_+(\xi)f(\xi) d\xi$. Furthermore, the driving term $f(\xi)$ is defined in equation \ref{fRHS}, and the normalized coefficients $\tilde{A}_\pm(\xi)$ are defined in \ref{modREF}.

The amplitudes of the incoming and outgoing components to $y_1(x)$ can now be clearly identified. Therefore, the modified reflection coefficient is given by the formula:
\begin{equation}
R = \left|\frac{r_0 + \epsilon y_{1,\text{out}}}{1 + \epsilon y_{1,\text{in}}} \right|^2=\left|\frac{r_0+\epsilon\left(\frac{2\pi ie^{-\pi\frac{\eta}{2}}}{\Gamma\left(-i\frac{\eta}{2} \right)\Gamma\left(1-i\frac{\eta}{2} \right)}\mathcal{F}^-_{-L}(0)-e^{-\pi\eta}\mathcal{A}^-_{-L}(0) \right)}{1-\epsilon \mathcal{F}^-_{-L}(0)} \right|^2
\label{modR}
\end{equation}

\noindent where $r_0 \doteq \tilde{A}e^{-\pi\eta} - \frac{2\pi i}{\Gamma\left(-i\frac{\eta}{2} \right)\Gamma\left(1- i \frac{\eta}{2} \right)}e^{-\pi \frac{\eta}{2}}$ is the unperturbed reflected wave amplitude. Note that $r_0$ may be $O(\epsilon)$ in the case of nearly-complete X-B mode-conversion; hence one should not expand R as a power series in $\epsilon$ for the general case.

\begin{figure}
\begin{overpic}[width=0.5\linewidth]{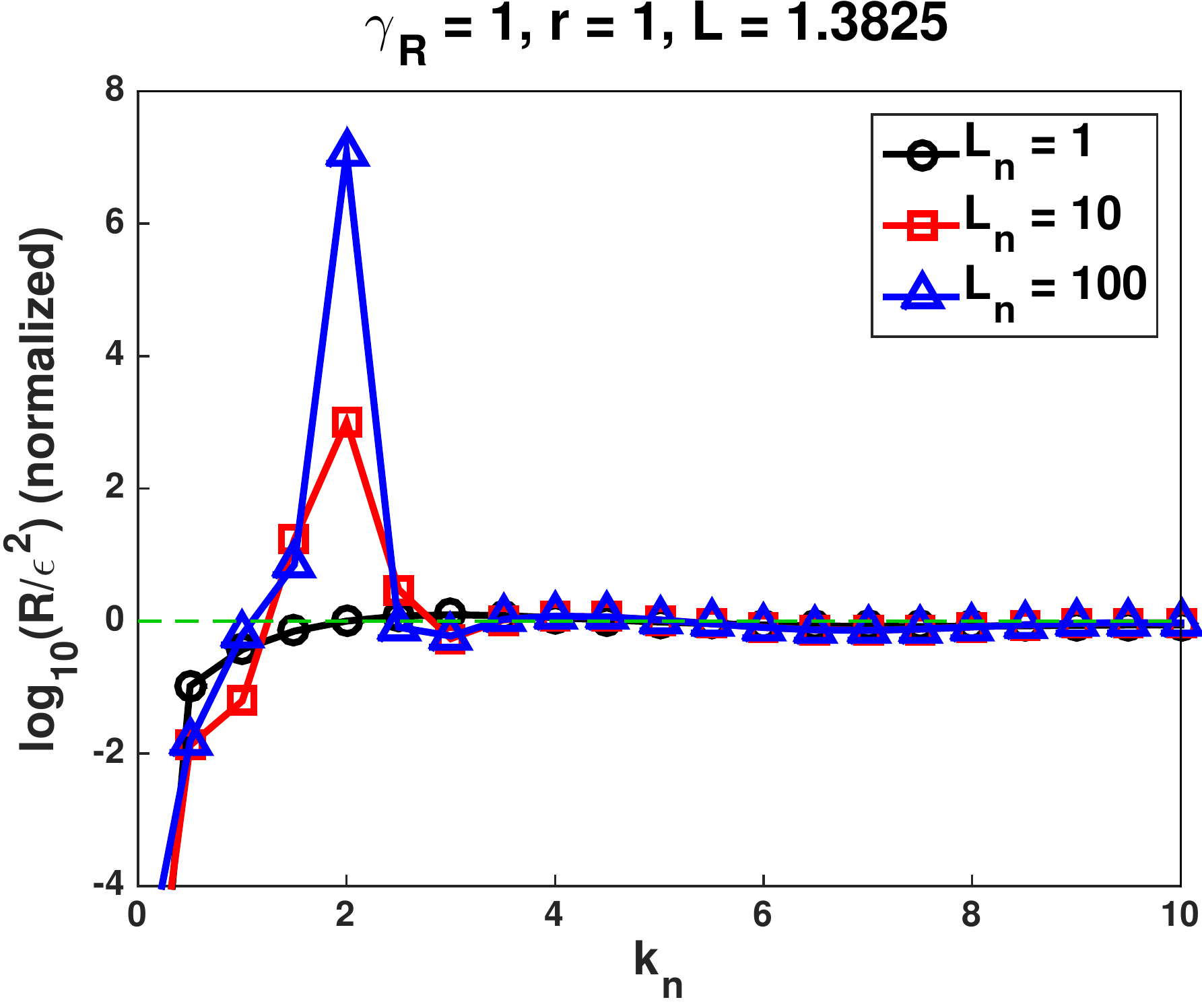}
\put(1,15){\large\textbf{(a)}}
\end{overpic}
\begin{overpic}[width=0.5\linewidth]{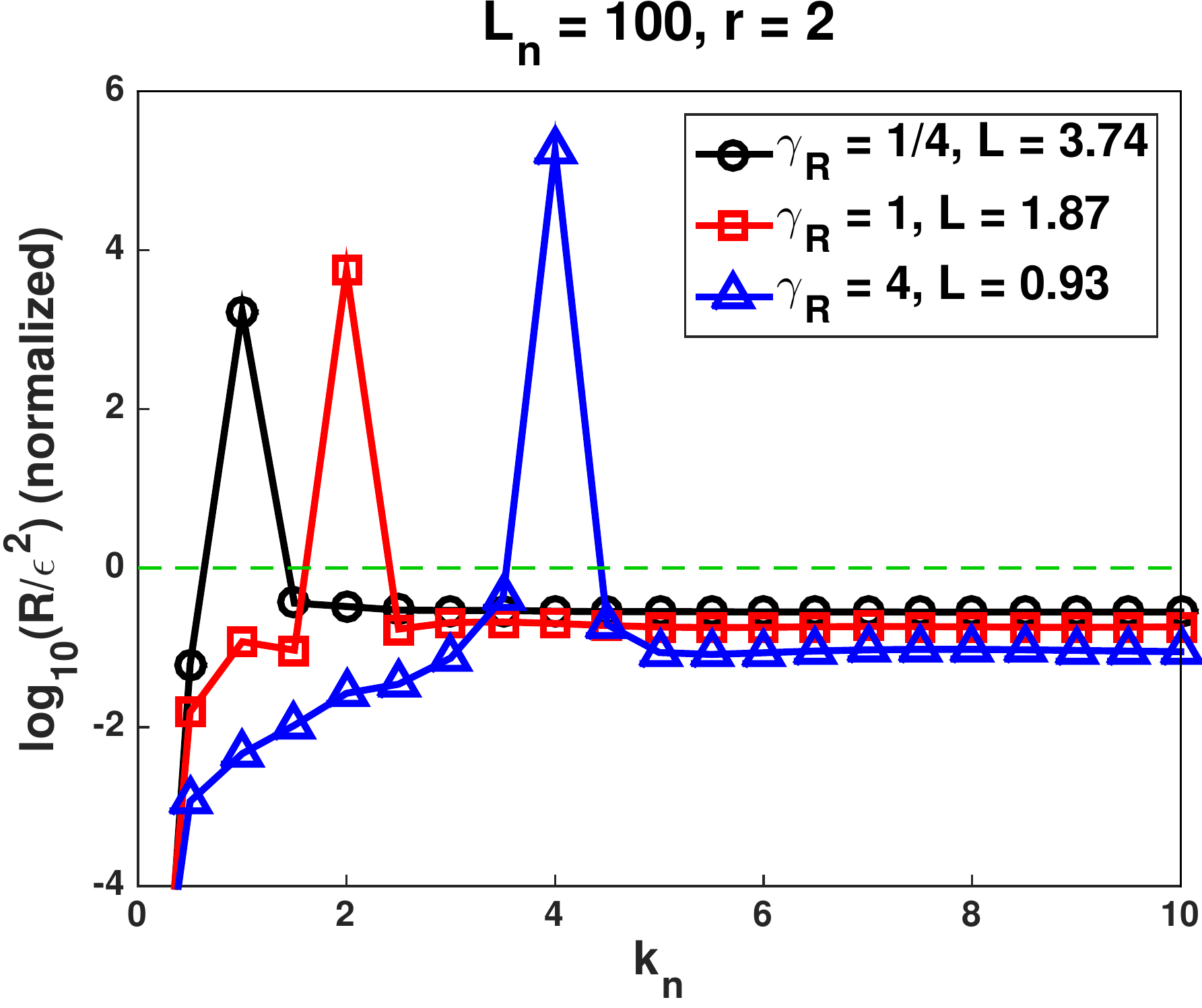}
\put(1,15){\large\textbf{(b)}}
\end{overpic}
\caption{The modified reflection coefficient normalized by the squared fluctuation amplitude $\epsilon^2$. \textbf{(a)} Variation in $R/\epsilon^2$ as the fluctuation wavenumber $k_n$ and the fluctuation decay length $L_n$ are varied, while the launched X-mode wavenumber $\sqrt{\gamma_R}$ is held constant, with no asymmetry ($r=1$) in the equilibrium. \textbf{(b)} Variation in $R/\epsilon^2$ as the fluctuation wavenumber $k_n$ and the launched X-mode wavenumber $\sqrt{\gamma_R}$ are varied, while the fluctuation decay length $L_n$ is held constant, with small asymmetry in the equilibrium. In both figures, all traces are presented on a log-scale, and normalized by their asymptotic values computed with equations \ref{AsymResultA} and \ref{AsymResultF}; for \textbf{(a)} this value is $\sim 0.71\cdot L_n^{-2}$, while for \textbf{(b)} these values are $5.89$, $0.15$, and $4.72\times 10^{-3}$, for $\gamma_R = 1/4$, $\gamma_R = 1$, and $\gamma_R = 4$ respectively. In all cases, the parameters are chosen such that if the density fluctuations were absent ($\epsilon = 0$), the X-B mode-conversion efficiency would be 100\% ($R=0$).}
\label{Bragg}
\end{figure}

In Figure \ref{Bragg}, the modified reflection coefficient (equation \ref{modR}) is numerically computed for select parameters. An adaptive Levin quadrature rule\cite{Levin96} is used to compute the highly-oscillatory integrals that constitute $\mathcal{F}^-_{-L}(0)$ and $\mathcal{A}^-_{-L}(0)$ using the Mathematica software\cite{Wolfram91}. Moreover, the Whittaker functions $W_k(z)$ are mapped onto the principal Riemann sheet $\text{Arg}(z) \in [-\frac{\pi}{2}, \frac{\pi}{2}]$ using the analytic continuation\cite{Heading62b}: 
\begin{equation}
W_k\left(|z|e^{-i\frac{3\pi}{2}}\right) = e^{-2\pi i k}W_k\left(|z|e^{i\frac{\pi}{2}} \right) - \frac{2\pi i e^{-i\pi k}}{\Gamma\left(1-k \right)\Gamma\left(-k \right)}W_{-k}\left(|z|e^{-i\frac{\pi}{2}} \right)
\end{equation}

\noindent In generating figure \ref{Bragg}, all the parameters are chosen such that $r_0 = 0$. In other words, the X-B mode-conversion would be complete in the absence of density fluctuations. For such parameters, as seen from equation \ref{modR}, $R$ is approximately quadratic in the fluctuation amplitude $\epsilon$, so $\frac{R}{\epsilon^2}$ will be nearly independent of the fluctuation amplitude. As a reminder, $\gamma_R = 1$ corresponds to an X-mode wave launched from vacuum, which is the relevant case for most X-B experiments. Other values of $\gamma_R$ are used in figure \ref{Bragg} to further illustrate the behavior of the modified reflection coefficient, since this behavior is not easily deduced from equation \ref{modR}.

In both plots of figure \ref{Bragg}, there is clear evidence of Bragg backscattering (BBS) in the modified X-B reflection coefficient. BBS occurs when the resonance condition $k_n = 2k(x) \approx 2\sqrt{\gamma_R}$ is satisfied, and constructive interference in the reflected wave pattern results. It was originally discovered in the context of x-ray crystallography\cite{Bragg13}, but has gained renewed interest in the hydrodynamics community through the design of Bragg breakwaters to protect shorelines\cite{Couston17}, and in the plasma physics community through studies on reflectometry and the O-X-B mode-conversion in the presence of turbulence\cite{Gusakov09,Popov15,Gospodchikov16}. Near the BBS resonance, the perturbative approach breaks down, and the reflection coefficient provided by equation \ref{modR} is not necessarily bounded by $1$. On the other hand, as seen in the first plot of figure \ref{Bragg}, the resonance response is reduced when the density fluctuations decay rapidly with distance (small $L_n$). This is because the spatial width of the Bragg resonance region is reduced for smaller $L_n$. It is interesting to note that only the fundamental Bragg resonance is visible in figure \ref{Bragg}. This is because (1) higher harmonic Bragg resonances are weaker than the fundamental resonance, and (2) we only consider X-mode waves that have no wavenumber component perpendicular to the direction of the background plasma inhomogeneity. When this restriction is lifted, such as when modelling a beam of finite width launched obliquely, the Bragg resonance should broaden into a Bragg spectrum due to the variation in wavenumber. Forward-scattering may also appear when higher-dimensional effects are considered, adding further structure to the modified reflection coefficient.

It should be noted that for most experimental parameters, fluctuation wavelengths are typically much larger than EC wavelengths, so the Bragg resonance is not expected to occur. However, resonant reflections may occur on devices that use EC waves to initiate plasma breakdown, since such methods can create EC-scale fluctuations, or on low-field devices such as STs, where EC waves have longer wavelengths than their high-field counterparts. On NSTX, for example, fluctuation wavelengths as small as a couple of centimeters ($\sim 6$~cm) have been observed in the edge plasma density\cite{Zweben15}. This is on the same order of magnitude as the EC wavelength on NSTX, which is approximately $2$~cm for a magnetic field strength of $0.5$~T.

We can approximate the modified reflection coefficient assuming the perturbation wavenumber is much larger than the vacuum wavenumber of the incident X-mode wave. Let $k_n \gg \sqrt{\gamma_R}$, $k_n \gg \frac{\sqrt{\gamma_R}}{r}$, and $k_n \gg L_n^{-1}$. Then, $\frac{\sin\left(k_n \xi \right)}{\xi}$ can be formally viewed as a nascent Dirac $\delta$-function. In this case, $\mathcal{A}^-_{-L}(0)$ and $\mathcal{F}^-_{-L}(0)$ are approximated as:
\numparts
\begin{eqnarray}
\label{AsymResultA}
\mathcal{A}^-_{-L}(0) \sim \frac{r^4\tilde{A}_-(0)+\tilde{A}_+(0)}{r^4+1}\mathcal{F}^-_{-L}(0)\\
\mathcal{F}^-_{-L}(0) \sim -2\pi\left(r^4+1 \right) \frac{\eta^2}{\gamma_R}\frac{\sqrt{2e}}{L_n}e^{-3\pi\frac{\eta}{4}}\left(\frac{\tilde{A}}{\Gamma\left(1+i\frac{\eta}{2} \right)} + \frac{1}{\Gamma\left(1-i\frac{\eta}{2} \right)}\right)
\label{AsymResultF}
\end{eqnarray}
\endnumparts

\noindent Hence, the reflection coefficient approaches a constant independent of $k_n$, and dependent on $L_n$ only via normalization. This asymptotic value of $R$ is shown in figure \ref{Bragg} as the dashed green line. All of the traces presented in this figure have been normalized by their respective asymptotic values such that they can all be viewed on the same scale. Moreover, as discussed in \ref{Perturb}, the asymptotic values of $\mathcal{A}^-_{-L}(0)$ and $\mathcal{F}^-_{-L}(0)$ can be used to assess the validity of the perturbative results.

\begin{figure}
\includegraphics[width=0.6\linewidth]{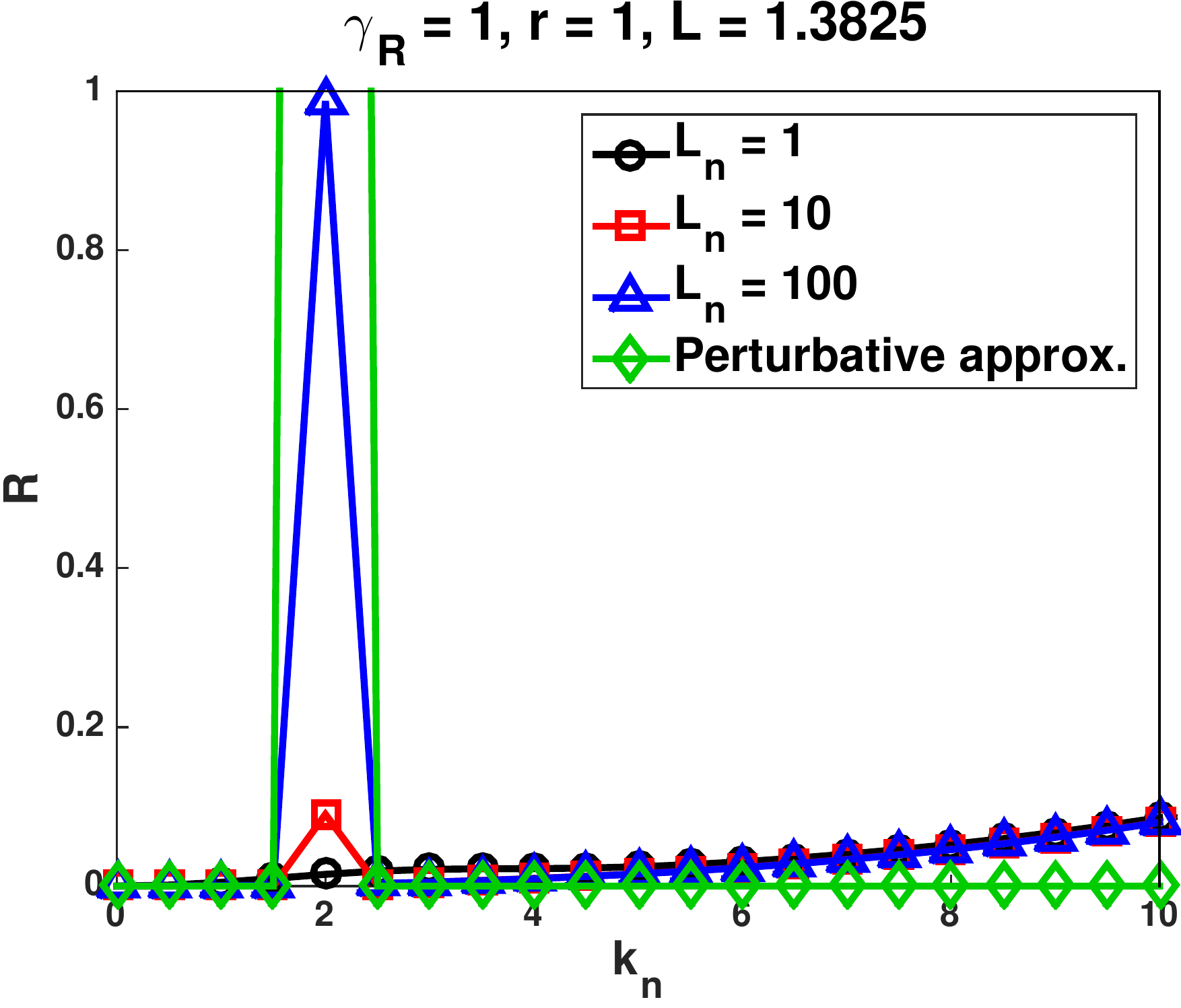}
\caption{A comparison between directly integrating the IB model equation when density fluctuations are present (equation \ref{fluctBUDDENeq}) and the approximate integral solution obtained with perturbation theory (equation \ref{modR}). The black, red, and blue traces show the modified reflection coefficients obtained by directly integrating equation \ref{fluctBUDDENeq} for various values of $L_n$, while the green trace shows the perturbative approximation to the modified reflection coefficient obtained from equation \ref{modR} with $L_n=100$. Here, the fluctuation amplitude is moderately large, with $\epsilon = 0.1$; however, at these parameters there is no creation of additional side-resonances.}
\label{BraggNUM}
\end{figure}

As shown in figure \ref{BraggNUM}, the analytical results are corroborated by the direct numerical integration of equation \ref{fluctBUDDENeq} with the IB boundary condition. In particular, one sees that the reflection coefficient is indeed bounded by $1$ in the vicinity of the Bragg resonance when the full form of the potential function is considered. The numerical integration is performed using an adaptive, error-controlled explicit Runge-Kutta (4,5) method (RK(4,5))\cite{Dormand80,Shampine97} on the domain $[-L,-\delta) \cup (\delta, x_c]$, where $x_c \gg 1$ is an arbitrary cutoff at large $x$ and $\delta \ll 1$ is an arbitrary cutoff around the pole at $x=0$. The amplitude absorption at the pole is put in by hand using the following interpolation procedure:
\begin{enumerate}
\numparts
\item{$E_y'(0)$ is computed using the Forward Euler integration rule:
\begin{equation}
E_y'(0^-) = E_y'(-\delta) - \delta\cdot\tilde{Q}(-\delta)E_y(-\delta)
\end{equation}}
\item{$E_y(0)$ is computed using the Trapezoid integration rule:
\begin{equation}
E_y(0) = E_y(-\delta) + \frac{\delta}{2}\left(E_y'(-\delta) + E_y'(0^-) \right)
\end{equation}}
\item{Flux loss is imposed via continuity of the derivative at $x=0$, noting the modified residual term due to the density fluctuations:
\begin{equation}
E_y'(0^+) = E_y'(0^-) - i\pi \frac{\beta \gamma_R^2(r^2+1)^2}{\gamma_R^2(r^2+1)^2-8\epsilon k_n \beta r^4} E_y(0)
\end{equation}}
\item{A predictor-corrector pair is used to interpolate the positive side of the pole. For the predictor stage, $\tilde{E}_y(\delta)$ is obtained via Forward Euler:
\begin{equation}
\tilde{E}_y(\delta) = E_y(0) + \delta\cdot E_y'(0^+)
\end{equation}}
\item{$\tilde{E}_y(\delta)$ is used to compute $E_y'(\delta)$ using the Backward Euler integration rule:
\begin{equation}
E_y'(\delta) = E_y'(0^+) - \delta\cdot \tilde{Q}(\delta)\tilde{E}_y(\delta)
\end{equation}}
\item{Finally, for the corrector stage, $E_y(\delta)$ is computed via Trapezoid rule:
\begin{equation}
E_y(\delta) = E_y(0) + \frac{\delta}{2}\left(E_y'(0^+) + E_y'(\delta) \right)
\end{equation}}
\endnumparts
\end{enumerate}

\noindent Since the interpolation near the pole uses a lower order integration scheme than the rest of the integration domain, it constitutes the dominant source of error.

It should be noted that in figure \ref{BraggNUM}, there is a small, but steady increase in the directly-integrated reflection coefficient outside of the resonance region. This is most likely a numerical artifact stemming from the increasingly fine-scale structure that is introduced by higher wavenumber perturbations. Indeed, increasing $k_n$ beyond about 20 causes $R$ to diverge as the adaptive algorithm used in the integration routine breaks down. This break-down occurs because the step size needed to resolve the potential function near the UHR when high wavenumber fluctuations are present is on the order of the machine precision. This leads us to remark on the two main shortcomings of the numerical integration procedure: (1) as just mentioned, the adaptive RK(4,5) method is unsuitable when fine structure becomes present in the potential function, either due to large $k_n$ or large $\epsilon$, and (2) the low-order treatment of the resonant absorption is unsuitable for the large $\epsilon$ regime when multiple resonances are present. It is anticipated that these shortcomings could be resolved with a more sophisticated integration scheme.

As a final remark, directly integrating equation \ref{fluctBUDDENeq} with the IB boundary condition is much faster than computing the approximate integral solutions given in equation \ref{modR} for the methods employed here. This is because numerically integrating the highly oscillatory terms of equation \ref{modR} requires a large number of recursion levels to achieve a reasonable accuracy, sometimes up to $30$. In fact, numerically integrating equation \ref{modR} takes on the order of minutes, while numerically integrating equation \ref{fluctBUDDENeq} takes on the order of tenths of seconds. This is not a particularly rigorous comparison, but it is provocative nonetheless. The time to integrate equation \ref{fluctBUDDENeq} can presumably be made even shorter by using an optimized scheme in a compiled language. This is encouraging for the prospects of the IB model as a computation tool.

\section{Conclusion}

In this work, we report on a new model for the X-B mode-conversion problem. This model replaces the high-field side left-hand cutoff of the X-mode wave by an infinite conducting barrier that reflects the X-mode wave field. Among its several advantages is the ease with which it can be numerically implemented; only a homogeneous Dirichlet boundary condition needs to be imposed at the left-hand cutoff, rather than a radiation boundary condition at infinity. As a result, the new boundary condition allows both solutions to the X-mode wave equation to be present within the mode-conversion region simultaneously, which improves the numerical stability. Although we focus on the mode-conversion process in one spatial dimension, the approach is readily amenable to higher spatial dimensions, allowing for the implementation of more realistic plasma equilibria. Importantly, the infinite barrier model is a useful tool which can be used by experimentalists for the rapid diagnosis of their observations.

We explicitly derive the Infinite Barrier model in 1-D, as this can be done analytically, and then compare the new model to an existing 1-D X-B model that was presented in \cite{Ram96}, known as the Double-cutoff model. Expectedly, there are quantitative differences between the two models; however, the qualitative behavior, such as the generic interference pattern and the generic periodicity with respect to the high-field side cutoff location, are successfully reproduced in the Infinite Barrier model. Moreover, the Infinite Barrier has an additional degree of freedom compared to the Double-cutoff model. This additional degree of freedom can be used to either bring the Infinite Barrier model into very close quantitative agreement with the Double-cutoff model, or to generate novel behavior of the X-B mode-conversion efficiency. Such flexibility is useful when attempting to fit to experimental data.

Often in spherical tokamaks, the X-B mode-conversion region is located in the plasma edge, where micro-turbulence and blobs can cause very abrupt cutoffs of the launched X-mode wave. This is problematic, as the interference pattern of the X-mode wave field in the mode-conversion region is very sensitive to the distance between the upper hybrid resonance and the left-hand cutoff. The Infinite Barrier model is well-suited to study this issue of edge fluctuations, since the abrupt cutoff behavior is included in the model by construction. The spatial and temporal demands on experimental measurements are thereby relaxed significantly when fitting the Infinite Barrier model to an experiment hampered with edge fluctuations, since exact details of the high-field side cutoff need not be resolved.

To demonstrate this feature, the Infinite Barrier model is used to estimate the effect of small-amplitude density fluctuations on the X-B mode-conversion efficiency in 1-D. This study is performed using a perturbative approach with a Green's function. We observe that reflections of the incident X-mode wave off the density fluctuations modifies the X-B mode-conversion efficiency in a complicated, but usually small manner. However, when the periodicity of the density fluctuations is half the wavelength of the launched X-mode wave, resonant Bragg backscattering of the X-mode wave may occur, significantly reducing the X-B mode-conversion efficiency. Although the analysis only considers quasi-monochromatic density perturbations, it is expected that broad-spectrum density fluctuations will also yield a reduction in the X-B mode-conversion efficiency. This will be the subject of further investigation.

The analytical results are confirmed by direct numerical integration of the Budden differential equation with additional density fluctuations present. The integration is performed with an adaptive Runge-Kutta method, while the resonant absorption at the pole is put in by hand as part of a predictor-corrector Euler-Trapezoid scheme used in the vicinity of the resonance. This numerical method works well in the limited parameter regime where both the fluctuation amplitude and wavenumber are small; however, when the fluctuations are larger, fine-scale structures such as side-resonances and additional cutoffs appear in the potential function, which cause the present integration routine to fail. A future computational study will explore alternative integration schemes to address this shortcoming as part of the development of a proof-of-principle numerical code to study the X-B mode-conversion efficiency in a realistic tokamak equilibrium.

\ack
Xin Zhang is kindly acknowledged for helpful conversations and assistance in generating some of the figures. AKR is supported by the U.S. Department of Energy Grant Nos. DE-FG02-91ER-54109, DE-FG02-99ER-54525-NSTX, and DE-SC0018090.

\appendix
\section{Limits and asymptotics of the Whittaker functions}

\label{WhitAsym}

In developing the linear system that governs the X-B mode-conversion problem, the limit of the Whittaker function and its derivative are needed as the argument of the Whittaker function, denoted as $z$, tends to $0$. For noninteger $k$, these limits are respectively\cite{Olver10}:
\begin{eqnarray}
\lim_{z\to0}W_k(z) &= -\frac{1}{k\Gamma\left(-k\right)}\\
\lim_{z\to0}W'_k(z) &= \frac{1}{\Gamma\left(-k\right)}\log(z) + \frac{\Psi\left(1-k\right)-2\Psi\left(1\right)}{\Gamma\left(-k\right)}
\end{eqnarray}

\noindent where $\Gamma\left(z\right)$ is the Gamma function, and $\Psi\left( z\right)\doteq \frac{\text{d}}{\text{d}z}\log\Gamma\left(z \right)$ is the digamma function\cite{Abramowitz70}. 

To determine the ratio of the incoming and outgoing wave amplitudes, the asymptotic representations of the Whittaker functions are also needed. These asymptotic representations depend on the argument of $z$. If $|\text{arg}(z)| \le \pi$\cite{Whittaker35}:
\begin{equation}
W_{k,m}(z) \sim z^k e^{-\frac{z}{2}}
\end{equation}

\noindent while if $-2\pi < \text{arg}(z) < -\pi$\cite{Heading62b}:
\begin{equation}
W_{k,m}(z) \sim z^k e^{-\frac{z}{2}} - \frac{2\pi i e^{-2\pi i k}}{\Gamma\left(\frac{1}{2}-m-k \right) \Gamma\left(\frac{1}{2}+m-k \right)} z^{-k}e^{\frac{z}{2}}
\end{equation}

For the Budden potential, a branch cut is placed in the bottom-half of the complex $x$ plane, and begins at $x=0$. Hence, the argument of $-x$ is $i\pi$. In this case, the asymptotic representations of the Whittaker functions are as follows. For $x\to\infty$
\begin{eqnarray}
W_{-i\frac{\eta}{2},\frac{1}{2}}\left(2\sqrt{\gamma_R}xe^{-i\frac{\pi}{2}}\right) \sim \left|2\sqrt{\gamma_R}x \right|^{-i\frac{\eta}{2}}e^{-\pi\frac{\eta}{4}}e^{i\sqrt{\gamma_R}x}\\
W_{i\frac{\eta}{2},\frac{1}{2}}\left(2\sqrt{\gamma_R}xe^{-i\frac{3\pi}{2}}\right) \sim \left|2\sqrt{\gamma_R}x \right|^{i\frac{\eta}{2}}e^{3\pi\frac{\eta}{4}}e^{-i\sqrt{\gamma_R}x}\nonumber\\
\hspace{3cm}- \left|2\sqrt{\gamma_R}x \right|^{-i\frac{\eta}{2}}\frac{2\pi ie^{\pi\frac{\eta}{4}}}{\Gamma\left(-i\frac{\eta}{2} \right)\Gamma\left(1-i\frac{\eta}{2} \right)} e^{i\sqrt{\gamma_R}x}
\end{eqnarray}

\noindent while for $x\to-\infty$
\begin{eqnarray}
W_{-i\frac{\eta r}{2},\frac{1}{2}}\left(2\frac{\sqrt{\gamma_R}}{r}x^{-i\frac{\pi}{2}}\right) \sim \left|2\frac{\sqrt{\gamma_R}}{r}x \right|^{-i\frac{\eta r}{2}}e^{\pi\frac{\eta r}{4}}e^{i\frac{\sqrt{\gamma_R}}{r}x}\\
W_{i\frac{\eta r}{2},\frac{1}{2}}\left(2i\frac{\sqrt{\gamma_R}}{r}x^{-i\frac{3\pi}{2}}\right) \sim \left|2\frac{\sqrt{\gamma_R}}{r}x \right|^{i\frac{\eta r}{2}}e^{\pi\frac{\eta r}{4}}e^{-i\frac{\sqrt{\gamma_R}}{r}x}
\end{eqnarray}

\noindent These asymptotic representations have the following physical interpretation: $W_{-i\frac{\eta r}{2},\frac{1}{2}}\left(2\frac{\sqrt{\gamma_R}}{r}x^{-i\frac{\pi}{2}}\right)$ represents a wave incident upon the UHR from the HFS that is perfectly transmitted; on the other hand, $W_{i\frac{\eta}{2},\frac{1}{2}}\left(2\sqrt{\gamma_R}x^{-i\frac{3\pi}{2}}\right)$ represents a wave incident upon the UHR from the LFS that is partially transmitted and partially reflected.

\section{Solving the DC and IB model equations for $\phi_\text{DC}$ and $\phi_\text{IB}$}
\label{PHIder}

For the DC model, the three boundary conditions to impose are the two continuity conditions given in equations \ref{mCond1} and \ref{mCond2}, and the radiation condition $C=0$. Using the limits of the Whittaker functions discussed in \ref{WhitAsym}, these become: 
\begin{eqnarray}
\frac{A}{\Gamma\left(i\frac{\eta_R}{2} \right)} + i\frac{\eta_R}{\eta_L}\frac{D}{\Gamma\left(-\frac{\eta_L}{2} \right)} = \frac{B}{\Gamma\left(-i\frac{\eta_R}{2} \right)}\\
i\frac{\Psi\left(1+i\frac{\eta_R}{2} \right)-\Psi\left(1-i\frac{\eta_R}{2} \right)+i\pi}{\Gamma\left(i\frac{\eta_R}{2} \right)}A =\nonumber\\
\frac{\log\left(\frac{\eta_R}{\eta_L} \right) +\Psi\left(1-\frac{\eta_L}{2} \right)-\Psi\left(1-i\frac{\eta_R}{2} \right)+i\frac{\pi}{2}}{\Gamma\left(-\frac{\eta_L}{2} \right)}\frac{\eta_R}{\eta_L}D
\end{eqnarray}

\noindent Solving for the ratio $\frac{A}{B}$, denoted as $\tilde{A}$, yields:

\begin{equation}
\tilde{A} = e^{-2i\theta_R}\frac{\log\left(\frac{\eta_R}{\eta_L} \right) + \Psi\left(1-\frac{\eta_L}{2} \right) - \Psi\left(1-i\frac{\eta_R}{2} \right) +i\frac{\pi}{2}}{\log\left(\frac{\eta_R}{\eta_L} \right) + \Psi\left(1-\frac{\eta_L}{2} \right) - \Psi\left(1+i\frac{\eta_R}{2} \right) -i\frac{\pi}{2}}
\label{AtildDC}
\end{equation}

\noindent where we have defined $\theta_R \doteq \text{Arg}\left(\Gamma\left(-i\frac{\eta}{2} \right) \right)$ to be the argument of the Gamma function $\Gamma\left(-i\frac{\eta}{2} \right)$. Clearly, $\tilde{A}$ is of unit modulus. This means that it is possible to express $\tilde{A} = e^{2i\phi_\text{DC}-2i\theta_R-i\pi}$ for some $\phi_\text{DC}$. By direct computation, $\phi_\text{DC}$ is obtained as:
\begin{eqnarray}
\phi_\text{DC} =\tan^{-1}\left(\frac{\log\left(\frac{\eta_R}{\eta_L}\right)+\Psi\left( 1-\frac{\eta_L}{2}\right)-\text{Re}\left[\Psi\left( 1-i\frac{\eta_R}{2}\right)\right]}{\text{Im}\left[\Psi\left( 1-i\frac{\eta_R}{2}\right)\right]-\frac{\pi}{2}} \right)
\end{eqnarray}

Finally, using this representation of $\tilde{A}$ and the asymptotic representations of the Whittaker functions, presented in \ref{WhitAsym}, the reflection coefficient for an X-mode wave incident from the LFS is calculated to be:
\begin{eqnarray}
R = \left|\tilde{A}e^{-\pi\eta} - \frac{2\pi i}{\Gamma\left(-i\frac{\eta}{2} \right)\Gamma\left(1- i \frac{\eta}{2} \right)}e^{-\pi \frac{\eta}{2}} \right|^2\nonumber\\
~~=1-4e^{-\pi\eta}\left(1-e^{-\pi\eta} \right)\cos^2\left(\phi_\text{DC} \right)
\label{RefCoef}
\end{eqnarray}

For the IB model, the three boundary conditions to impose are the two continuity conditions (\ref{mCond1} and \ref{mCond2}) and the IB boundary condition $y_L(-L) = 0$. Let us define the normalized quantities $\tilde{A} \doteq \frac{A}{B}$, $\tilde{C} \doteq \frac{C}{B}$, and $\tilde{D} \doteq \frac{D}{B}$. Then, the three boundary conditions equations constitute a linear system in $\tilde{A}$, $\tilde{C}$, and $\tilde{D}$:
\begin{equation}
\hspace{-1.4cm}\left(\begin{array}{c}
\tilde{A}\\[3mm]
\tilde{C}\\[3mm]
\tilde{D}
\end{array}\right)^\textsf{\small T}
\left(\begin{array}{c c c}
\frac{1}{\Gamma\left(i\frac{\eta}{2} \right)} & \frac{\Psi\left(1+i\frac{\eta}{2}\right)-\Psi\left(1-i\frac{\eta}{2}\right)+i\pi}{\Gamma\left(i\frac{\eta}{2} \right)} & 0\\[3mm]
-\frac{1}{r\Gamma\left(i\frac{\eta r}{2} \right)} & \frac{\log(r)+\Psi\left(1-i\frac{\eta}{2}\right)-\Psi\left(1+i\frac{\eta r}{2}\right)-i\pi}{r\Gamma\left(i\frac{\eta r}{2} \right)} & e^{-2i\theta_W}\\[3mm]
\frac{1}{r\Gamma\left(-i\frac{\eta r}{2} \right)} & \frac{\Psi\left(1-i\frac{\eta r}{2}\right)-\Psi\left(1-i\frac{\eta}{2}\right)-\log(r)}{r\Gamma\left(-i\frac{\eta r}{2} \right)} & 1
\end{array}\right)
=\left(\begin{array}{c}
\frac{1}{\Gamma\left(-i\frac{\eta}{2} \right)}\\[3mm]
0\\[3mm]
0
\end{array}\right)^\textsf{\small T}
\label{linSys}
\end{equation}
\noindent where we have defined $\theta_W \doteq \text{Arg}\left(W_{i\frac{\eta r}{2}}\left(2\sqrt{\gamma_L}L e^{-i\frac{\pi}{2}}\right) \right)$, and $\intercal$ denotes the transpose of the column vector. The first two columns of \ref{linSys} correspond to the two continuity conditions, while the final column corresponds to the IB condition. System \ref{linSys} is readily inverted to yield $\tilde{A}$, $\tilde{C}$, and $\tilde{D}$. In particular:
\begin{equation}
\tilde{A} = e^{-2i\left(\theta_R+\theta_L+\theta_W\right)}
\frac{\left(\Upsilon +\frac{2i}{\eta r}-i\pi\left[\coth\left(\pi\frac{\eta r}{2}\right)+1 \right] \right)e^{2i\left(\theta_L+\theta_W\right)} + \Upsilon}{\left(\Upsilon \strut^*-\frac{2i}{\eta r}+i\pi\left[\coth\left(\pi\frac{\eta r}{2}\right)+1 \right] \right)e^{-2i\left(\theta_L+\theta_W\right)} + \Upsilon \strut^*}
\end{equation}

\noindent where we have defined $\theta_L \doteq \text{Arg}\left(\Gamma\left(-i\frac{\eta r}{2} \right) \right)$ and $\Upsilon \doteq \Psi\left(1-i\frac{\eta}{2} \right)-\Psi\left(1-i\frac{\eta r}{2} \right)+\log(r)$.

As with the DC model, $\tilde{A}$ is of unit modulus. Expressing $\tilde{A} = e^{2i\phi_\text{IB}-2i\theta_R-i\pi}$ for some $\phi_\text{IB}$ yields:
\begin{eqnarray}
\hspace{-2.6cm}\phi_\text{IB} = \tan^{-1}\left(\frac{\sin\left(2\theta_L + 2\theta_W \right)\left[\pi + \pi\coth\left(\pi\frac{\eta r}{2} \right) -\Upsilon_\text{I} - \frac{2}{\eta r}\right]+\Upsilon_\text{R}\left[\cos\left(2\theta_L + 2\theta_W \right)+1\right]}{\cos\left(2\theta_L + 2\theta_W \right)\left[\pi + \pi\coth\left(\pi\frac{\eta r}{2} \right) -\Upsilon_\text{I} - \frac{2}{\eta r}\right]-\Upsilon_\text{R}\sin\left(2\theta_L + 2\theta_W \right) - \Upsilon_\text{I}} \right)\nonumber\\
\hspace{-1.5cm}- \theta_L - \theta_W 
\label{IBphase}
\end{eqnarray}

\noindent where $\Upsilon_\text{R}$ and $\Upsilon_\text{I}$ denote the real and imaginary parts of $\Upsilon$ respectively. As before, the unitarity of $\tilde{A}$ implies that the reflection coefficient for the IB model is also given by equation \ref{RefCoef}, albeit with the phase specified by equation \ref{IBphase}.

\section{The $\textbf{M}$ matrices  and $\textbf{b}$ vectors for computing the IB model Green's function}
\label{IBGreen}

The matrix $\textbf{M}_{(+)}$, and the vector $\textbf{b}_+$ that appear in equation \ref{linPLUS} are given as:
\begin{eqnarray}
\textbf{M}_{(+)}= \nonumber\\
\hspace{-2.8cm}\left(\begin{array}{c c c c c}
0 & W_{-i\frac{\eta}{2}}\left(\varsigma e^{-i\frac{\pi}{2}} \right) & 0 & W'_{-i\frac{\eta}{2}}\left(\varsigma e^{-i\frac{\pi}{2}} \right) & 0\\[2mm]
\frac{1}{\Gamma\left(i\frac{\eta}{2} \right)} & -W_{-i\frac{\eta}{2}}\left(\varsigma e^{-i\frac{\pi}{2}} \right) & \frac{\Psi\left(1+i\frac{\eta}{2}\right)-\Psi\left(1-i\frac{\eta}{2}\right)+i\pi}{\Gamma\left(i\frac{\eta}{2} \right)} & -W'_{-i\frac{\eta}{2}}\left(\varsigma e^{-i\frac{\pi}{2}} \right) & 0\\[2mm]
-\frac{1}{\Gamma\left(-i\frac{\eta}{2} \right)} & -W_{i\frac{\eta}{2}}\left(\varsigma e^{-i\frac{3\pi}{2}} \right) & 0 & W'_{i\frac{\eta}{2}}\left(\varsigma e^{-i\frac{3\pi}{2}} \right) & 0\\[2mm]
-\frac{1}{r\Gamma\left(i\frac{\eta r}{2} \right)} & 0 & \frac{\log(r)+\Psi\left(1-i\frac{\eta}{2}\right)-\Psi\left(1+i\frac{\eta r}{2}\right)-i\pi}{r\Gamma\left(i\frac{\eta r}{2} \right)} & 0 & e^{-2i\theta_W}\\[2mm]
\frac{1}{r\Gamma\left(-i\frac{\eta r}{2} \right)} & 0 & \frac{\Psi\left(1-i\frac{\eta r}{2}\right)-\Psi\left(1-i\frac{\eta}{2}\right)-\log(r)}{r\Gamma\left(-i\frac{\eta r}{2} \right)} & 0 & 1
\end{array}\right)\\
\textbf{b}_+ \doteq \left(\begin{array}{c c c c c}
0, & -W_{i\frac{\eta}{2}}\left(\varsigma e^{-i\frac{3\pi}{2}} \right), & 0, & W'_{i\frac{\eta}{2}}\left(\varsigma e^{-i\frac{3\pi}{2}} \right) + \frac{i}{2\sqrt{\gamma_R}B_+}, & 0
\end{array}\right)
\end{eqnarray}

\noindent The matrix $\textbf{M}_{(-)}$, and the vector $\textbf{b}_-$ that appear in equation \ref{linMIN} are given as:
\begin{eqnarray}
\textbf{M}_{(-)}=\nonumber\\
\hspace{-2.8cm}\left(\begin{array}{c c c c c}
\frac{1}{\Gamma\left(i\frac{\eta}{2} \right)} & 0 & \frac{\Psi\left(1+i\frac{\eta}{2}\right)-\Psi\left(1-i\frac{\eta}{2}\right)+i\pi}{\Gamma\left(i\frac{\eta}{2} \right)} & 0 & 0\\[2mm]
-\frac{1}{r\Gamma\left(i\frac{\eta r}{2} \right)} & W_{-i\frac{\eta r}{2}}\left(\frac{\varsigma}{r}e^{i\frac{\pi}{2}} \right) & \frac{\log(r)+\Psi\left(1-i\frac{\eta}{2}\right)-\Psi\left(1+i\frac{\eta r}{2}\right)-i\pi}{r\Gamma\left(i\frac{\eta r}{2} \right)} & W'_{-i\frac{\eta r}{2}}\left(\frac{\varsigma}{r}e^{i\frac{\pi}{2}} \right) & 0\\[2mm]
\frac{1}{r\Gamma\left(-i\frac{\eta r}{2} \right)} & W_{i\frac{\eta r}{2}}\left(\frac{\varsigma}{r}e^{-i\frac{\pi}{2}} \right) & \frac{\Psi\left(1-i\frac{\eta r}{2}\right)-\Psi\left(1-i\frac{\eta}{2}\right)-\log(r)}{r\Gamma\left(-i\frac{\eta r}{2} \right)} & -W'_{i\frac{\eta r}{2}}\left(\frac{\varsigma}{r}e^{-i\frac{\pi}{2}} \right) & 0\\[2mm]
0 & -W_{-i\frac{\eta r}{2}}\left(\frac{\varsigma}{r}e^{i\frac{\pi}{2}} \right) & 0 & -W'_{-i\frac{\eta r}{2}}\left(\frac{\varsigma}{r}e^{i\frac{\pi}{2}} \right) & e^{-2i\theta_W}\\[2mm]
0 & -W_{i\frac{\eta r}{2}}\left(\frac{\varsigma}{r}e^{-i\frac{\pi}{2}} \right) & 0 & W'_{i\frac{\eta r}{2}}\left(\frac{\varsigma}{r}e^{-i\frac{\pi}{2}} \right) & 1
\end{array}\right)\\
\textbf{b}_- \doteq \left(\begin{array}{c c c c c}
\frac{1}{\Gamma\left(-i\frac{\eta}{2} \right)}, & 0, & 0, & \frac{ir}{2\sqrt{\gamma_R}B_-}, & 0
\end{array} \right)
\end{eqnarray}

\noindent Finally, the matrix $\textbf{M}_{(0)}$, and the vector $\textbf{b}_0$ that appear in equation \ref{linZERO} are given as:

\begin{eqnarray}
\textbf{M}_{(0)} =
\left(\begin{array}{c c c}
\frac{1}{\Gamma\left(i\frac{\eta}{2} \right)} & \frac{\Psi\left(1+i\frac{\eta}{2}\right)-\Psi\left(1-i\frac{\eta}{2}\right)+i\pi}{\Gamma\left(i\frac{\eta}{2} \right)} & 0\\[3mm]
-\frac{1}{r\Gamma\left(i\frac{\eta r}{2} \right)} & \frac{\log(r)+\Psi\left(1-i\frac{\eta}{2}\right)-\Psi\left(1+i\frac{\eta r}{2}\right)-i\pi}{r\Gamma\left(i\frac{\eta r}{2} \right)} & e^{-2i\theta_W}\\[3mm]
\frac{1}{r\Gamma\left(-i\frac{\eta r}{2} \right)} & \frac{\Psi\left(1-i\frac{\eta r}{2}\right)-\Psi\left(1-i\frac{\eta}{2}\right)-\log(r)}{r\Gamma\left(-i\frac{\eta r}{2} \right)} & 1
\end{array}\right)\\
\textbf{b}_0 \doteq \left(\begin{array}{c c c}
\frac{1}{\Gamma\left(-i\frac{\eta}{2} \right)}, & \frac{i}{2\sqrt{\gamma_R} B_0}, & 0
\end{array}\right)
\end{eqnarray}

\noindent In the above expressions, we have defined the quantity $\varsigma \doteq 2\sqrt{\gamma_R} |\xi|$. Also, the derivative of the Whittaker function is understood via the recurrence relation\cite{Olver10}:
\begin{equation}
W'_{k,\frac{1}{2}}(x) = \frac{(x-2k)W_{k,\frac{1}{2}}(x)-2W_{1+k,\frac{1}{2}}(x)}{2x}
\end{equation}

\noindent In principle, these linear systems can be inverted analytically; however the results are quite lengthy, and will not be presented here.

\section{The modified reflection coefficient for analytic density fluctuations which vanish at the UHR}
\label{modREF}

Let $\tilde{X}_+(\xi)$, $\tilde{X}_-(\xi)$, and $\tilde{X}_0(0)$, $X\in\{A,C,D,F,H \}$ denote the solutions to equations \ref{linPLUS}, \ref{linMIN}, \ref{linZERO} respectively. Since $\textbf{M}_{(+)}$, $\textbf{M}_{(-)}$, and $\textbf{M}_{(0)}$ are each non-singular, these solutions are well-defined. Then, $y_1(x)$ is given by the expression:
\begin{equation}
y_1(x) = \left\{\begin{array}{l l}
B_+\mathcal{I}_1(x) + B_+\mathcal{I}_2(x) + B_-\mathcal{I}_4(x) - i\pi B_0 \mathcal{R}_+(x) & x > 0\\
B_+\mathcal{I}_3(x) + B_-\mathcal{I}_5(x) + B_-\mathcal{I}_6(x) - i\pi B_0 \mathcal{R}_-(x) & x \le 0
\end{array}\right.
\end{equation}

\noindent where we have defined the integral quantities:
\begin{eqnarray}
I_1(x) = \int_{0}^x G_1(x,\xi)f(\xi)d\xi\\
I_2(x) = \int_{x}^\infty G_2(x,\xi)f(\xi)d\xi\\
I_3(x) = \int_{0}^\infty G_3(x,\xi)f(\xi)d\xi\\
I_4(x) = \int_{-L}^0 G_4(x,\xi)f(\xi)d\xi \\
I_5(x) = \int_{-L}^x G_5(x,\xi)f(\xi)d\xi\\
I_6(x) = \int_{x}^0 G_6(x,\xi)f(\xi)d\xi
\end{eqnarray}

\noindent we have defined the residue quantities:
\begin{eqnarray}
R_+(x) = \frac{16\beta^2 r^4}{\gamma_R^2\left(r^2+1\right)^2}\tilde{n}'(0)y_0(0)G_7(x,0) \\
R_-(x) = \frac{16\beta^2 r^4}{\gamma_R^2\left(r^2+1\right)^2}\tilde{n}'(0)y_0(0)G_8(x,0)
\end{eqnarray}

\noindent and we have defined the piecewise constituents of the Green's function:
\begin{eqnarray}
G_1(x,\xi) = \tilde{A}_+(\xi) W_{-i\frac{\eta}{2}}\left(2\sqrt{\gamma_R}x e^{-i\frac{\pi}{2}}\right) + W_{i\frac{\eta}{2}}\left(2\sqrt{\gamma_R}x e^{-i\frac{3\pi}{2}}\right)\\
G_2(x,\xi) = \tilde{C}_+(\xi) W_{-i\frac{\eta}{2}}\left(2\sqrt{\gamma_R}x e^{-i\frac{\pi}{2}}\right) + \tilde{D}_+(\xi) W_{i\frac{\eta}{2}}\left(2\sqrt{\gamma_R}x e^{-i\frac{3\pi}{2}}\right)\\
G_3(x,\xi) = \tilde{F}_+(\xi) W_{-i\frac{\eta r}{2}}\left(2\frac{\sqrt{\gamma_R}}{r}x e^{-i\frac{\pi}{2}}\right)\nonumber\\
\hspace{2cm}+\tilde{H}_+(\xi) W_{i\frac{\eta r}{2}}\left(2\frac{\sqrt{\gamma_R}}{r}x e^{-i\frac{3\pi}{2}}\right)\\
G_4(x,\xi) = \tilde{A}_-(\xi) W_{-i\frac{\eta}{2}}\left(2\sqrt{\gamma_R}x e^{-i\frac{\pi}{2}}\right) + W_{i\frac{\eta}{2}}\left(2\sqrt{\gamma_R}x e^{-i\frac{3\pi}{2}}\right)\\
G_5(x,\xi) = \tilde{C}_-(\xi) W_{-i\frac{\eta r}{2}}\left(2\frac{\sqrt{\gamma_R}}{r}x e^{-i\frac{\pi}{2}}\right)\nonumber\\
\hspace{2cm}+ \tilde{D}_-(\xi) W_{i\frac{\eta r}{2}}\left(2\frac{\sqrt{\gamma_R}}{r}x e^{-i\frac{3\pi}{2}}\right)\\
G_6(x,\xi) = \tilde{F}_-(\xi) W_{-i\frac{\eta r}{2}}\left(2\frac{\sqrt{\gamma_R}}{r}x e^{-i\frac{\pi}{2}}\right)\nonumber\\
\hspace{2cm}+ \tilde{H}_-(\xi) W_{i\frac{\eta r}{2}}\left(2\frac{\sqrt{\gamma_R}}{r}x e^{-i\frac{3\pi}{2}}\right)\\
G_7(x,0) = \tilde{A}_0(0) W_{-i\frac{\eta}{2}}\left(2\sqrt{\gamma_R}x e^{-i\frac{\pi}{2}}\right) + W_{i\frac{\eta}{2}}\left(2\sqrt{\gamma_R}x e^{-i\frac{3\pi}{2}}\right)\\
G_8(x,\xi) = \tilde{C}_0(0) W_{-i\frac{\eta r}{2}}\left(2\frac{\sqrt{\gamma_R}}{r}x e^{-i\frac{\pi}{2}}\right)\nonumber\\
\hspace{2cm}+ \tilde{D}_0(0) W_{i\frac{\eta r}{2}}\left(2\frac{\sqrt{\gamma_R}}{r}x e^{-i\frac{3\pi}{2}}\right)
\end{eqnarray}

\noindent It is understood that the singular integrals are evaluated according to their principal values. Moreover, $f(\xi)$ is given by equation \ref{fRHS}.

Let us now compute the modified X-B reflection coefficient in the presence of density fluctuations. As there is only a single boundary condition that the wave fields must satisfy, the remaining degree of freedom allows one to set $B_+ = B_- = B_0 = B = e^{-3\pi\frac{\eta}{4}}$ inconsequentially. This choice means that $y_0(x)$ is the $O(1)$ medium response to an incident wave of unit amplitude, and $y_1(x)$ is composed of the $O(\epsilon)$ unit-impulse responses of the medium for an incident wave of unit amplitude. 

To compute the modified reflection coefficient, the amplitudes of the outgoing and incoming components to $y_1(x)$ as $x \to \infty$ must be determined. This requires computing the asymptotics of 4 terms as $x \to \infty$: $R_+(x)$, $I_1(x)$, $I_2(x)$, and $I_3(x)$. Using the known asymptotics of the Whittaker functions, the asymptotics of each of these terms are given as:
\begin{eqnarray}
R_+(x)\sim \left(\frac{16\beta^2 r^4}{\gamma_R^2\left(r^2+1\right)^2}\tilde{n}'(0)y_0(0) e^{3\pi\frac{\eta}{4}}\right)\left(\frac{}{}y_-(x)\right.\nonumber\\
\hspace{1.3cm}\left.+\left[\tilde{A}_0(0)e^{-\pi\eta} - \frac{2\pi ie^{-\pi\frac{\eta}{2}}}{\Gamma\left(-i\frac{\eta}{2} \right)\Gamma\left(1-i\frac{\eta}{2} \right)} \right]y_+(x)\right)\\
I_1(x)\sim e^{3\pi\frac{\eta}{4}}\left(\frac{}{}\mathcal{F}^-_{\infty}(x)y_-(x)\right.\nonumber\\
\hspace{1.3cm}\left.+\left[e^{-\pi\eta}\mathcal{A}^-_{\infty}(x) - \frac{2\pi ie^{-\pi\frac{\eta}{2}}}{\Gamma\left(-i\frac{\eta}{2} \right)\Gamma\left(1-i\frac{\eta}{2} \right)} \mathcal{F}^-_\infty(x) \right]y_+(x)\right)\\
I_2(x)\sim e^{3\pi\frac{\eta}{4}}\left(\left[\int_{x}^\infty \tilde{D}_+(\xi)f(\xi) d\xi \right]y_-(x)\right.\nonumber\\
\hspace{1.3cm}\left.+ \left[e^{-\pi\eta}\int_{x}^\infty \tilde{C}_+(\xi)f(\xi)d\xi \right.\right.\nonumber\\
\hspace{1.3cm}\left.\left.- \frac{2\pi ie^{-\pi\frac{\eta}{2}}}{\Gamma\left(-i\frac{\eta}{2} \right)\Gamma\left(1-i\frac{\eta}{2} \right)} \int_{x}^\infty \tilde{D}_+(\xi)f(\xi)d\xi \right]y_+(x) \right)\\
I_4(x)\sim e^{3\pi\frac{\eta}{4}}\left(\left[\int_{-L}^0f(\xi) d\xi \right]y_-(x)\right.\nonumber\\
\hspace{1.3cm}\left.+ \left[e^{-\pi\eta}\int_{-L}^0 \tilde{A}_-(\xi)f(\xi)d\xi\right.\right.\nonumber\\
\hspace{1.3cm}\left.\left.- \frac{2\pi ie^{-\pi\frac{\eta}{2}}}{\Gamma\left(-i\frac{\eta}{2} \right)\Gamma\left(1-i\frac{\eta}{2} \right)} \int_{-L}^0f(\xi)d\xi \right]y_+(x) \right)\\
\end{eqnarray}

\noindent where we have defined the functions:
\begin{eqnarray}
y_+(x) \doteq |2\sqrt{\gamma_R}x |^{-i\frac{\eta}{4}}e^{i\sqrt{\gamma_R}x}\\
y_-(x) \doteq |2\sqrt{\gamma_R}x |^{i\frac{\eta}{4}}e^{-i\sqrt{\gamma_R}x}
\end{eqnarray}

\noindent to represent outgoing and incoming waves, respectively. For convenience, we have also introduced the following family of functions, defined for $x \ge 0$:
\begin{eqnarray}
\mathcal{A}^\pm_{s}(x) \doteq \int_0^s \tilde{A}_{\text{sgn}(s)}(\xi)f(\xi)d\xi \pm \int_{x}^\infty \tilde{A}_+(\xi)f(\xi)d\xi\\
\mathcal{F}^\pm_{s}(x) \doteq \int_0^s f(\xi) d\xi \pm \int_x^\infty f(\xi) d\xi
\end{eqnarray}

\noindent The modified reflection coefficient can then be constructed by identifying the coefficients of $y_+(x)$ and $y_-(x)$ respectively as $y_{1,\text{out}}$ and $y_{1,\text{in}}$ in equation \ref{modR}.

\section{Validity of the perturbative approach}
\label{Perturb}

\begin{figure}
\includegraphics[scale=0.8]{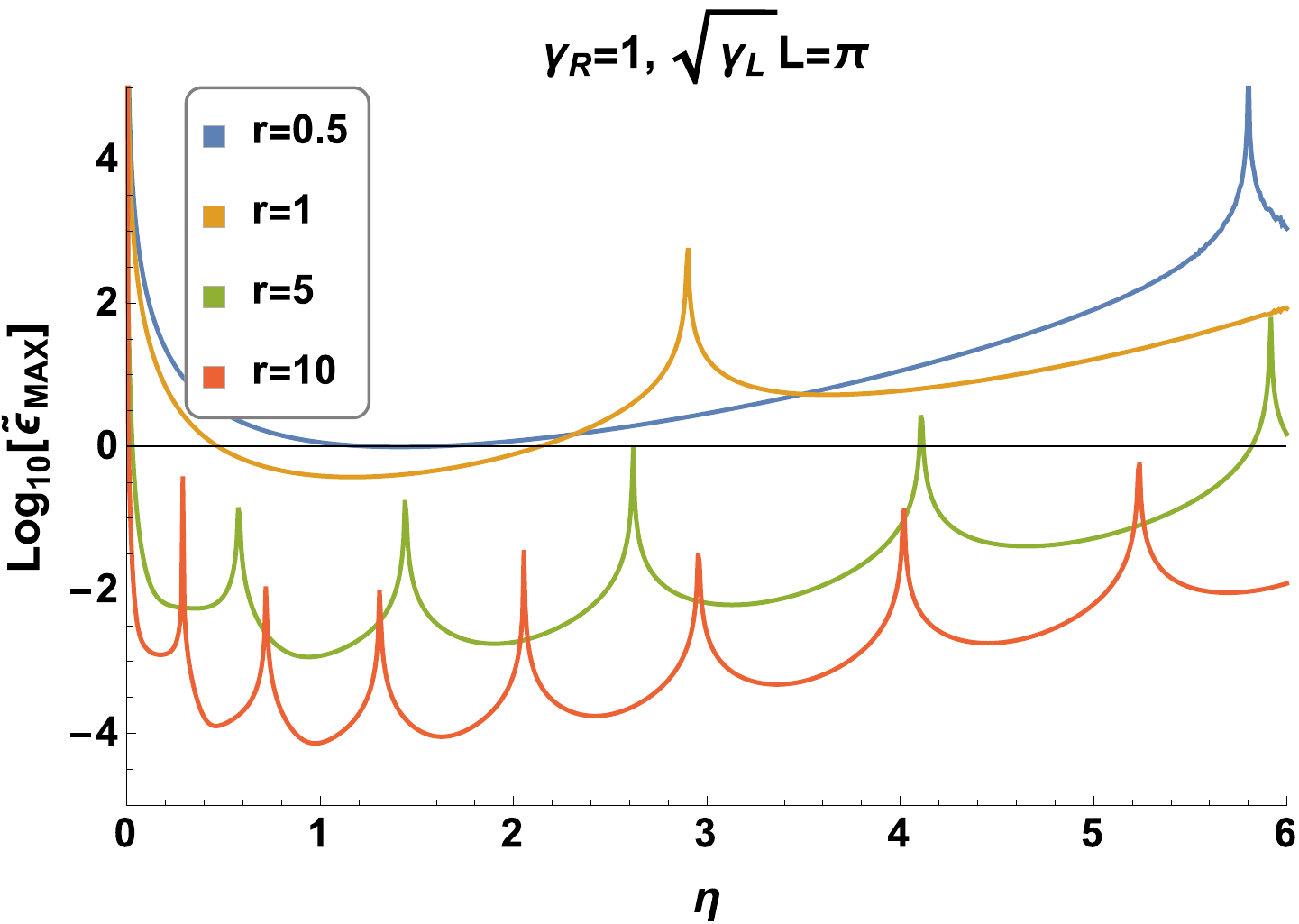}
\caption{Estimated maximum fluctuation amplitude for equation \ref{modR} to be valid. Typically, increasing $r$, decreasing $\eta$, and decreasing $\gamma_R$ (not shown) causes $\tilde{\epsilon}_\text{max}$ to decrease. Changing $L$ (not shown) only slightly modifies $\tilde{\epsilon}_\text{max}$. Note that $\tilde{\epsilon}_\text{max}$ is presented on a log-scale.}
\label{epsMAX}
\end{figure}

The perturbative approach used in the previous subsections is only valid when $\epsilon$ is `small', in a sense that was left unspecified. It is difficult to place a tight upper bound on the fluctuation amplitude $\epsilon$, beyond which the perturbative expansion of $y(x)$ would break down uniformly. An intuitive approach would be to identify the values of $\epsilon$ marking the appearance of additional SRs in the modified potential for various parameters; however, this approach would involve counting the roots to a sequence of nonlinear equations, which is not practical.

Instead, an approximate upper bound can be placed via the following argument. For the perturbative expansion of $y(x)$ to be valid, it is necessary that the modified reflection coefficient is always less than or equal to unity. Furthermore, for $R \le 1$, it is certainly necessary that:
\begin{equation}
\epsilon \left|\frac{2\pi ie^{-\pi\frac{\eta}{2}}}{\Gamma\left(-i\frac{\eta}{2} \right)\Gamma\left(1-i\frac{\eta}{2} \right)} \mathcal{F}^-_{-L}(0) - e^{-\pi\eta}\mathcal{A}^-_{-L}(0)\right| \le 1
\end{equation}

\noindent for else $R$ could be larger than one whenever $r_0 = 0$. This inequality must also be true when $\mathcal{A}^-_{-L}(0)$ and $\mathcal{F}^-_{-L}(0)$ take their asymptotic forms. 

Thus, an approximate upper bound on $\epsilon$ is arrived at:
\begin{equation}
\epsilon \le \frac{L_n}{\sqrt{2e}}\tilde{\epsilon}_\text{max} \doteq \left|\frac{2\pi ie^{-\pi\frac{\eta}{2}}}{\Gamma\left(-i\frac{\eta}{2} \right)\Gamma\left(1-i\frac{\eta}{2} \right)} \mathcal{F}^-_{-L}(0) - e^{-\pi\eta}\mathcal{A}^-_{-L}(0)\right|^{-1}
\end{equation}

\noindent where $\mathcal{A}^-_{-L}(0)$ and $\mathcal{F}^-_{-L}(0)$ are evaluated according to equations \ref{AsymResultA} and \ref{AsymResultF}. The upper bound $\tilde{\epsilon}_\text{max}$ only depends on the parameters of the wave field and the equilibrium ($\sqrt{\gamma_R}$, $\eta$, $r$, and $L$), and is independent of the details of the density fluctuations. The normalization $\frac{\sqrt{2e}}{L_n}$ has been explicitly factored out of $\tilde{\epsilon}_\text{max}$.

A coarse parameter scan reveals three general trends in the value of $\tilde{\epsilon}_\text{max}$: (1) $\tilde{\epsilon}_\text{max}$ tends to increase for increasing values of $\sqrt{\gamma_R}$, (2) $\tilde{\epsilon}_\text{max}$ tends to increase for increasing values of $\eta$, and (3) $\tilde{\epsilon}_\text{max}$ tends to decrease for increasing values of $r$. The dependence of $\tilde{\epsilon}_\text{max}$ on $L$ appears to be weaker than the other three parameters. These trends are understood concisely by the statement that $\tilde{\epsilon}_\text{max}$ is positively correlated with the height of the unperturbed Budden potential function. As can be seen in figure \ref{epsMAX}, large asymmetry can cause $\tilde{\epsilon}_\text{max}$ to be as low as $10^{-4}$; however for moderate values of the asymmetry parameter, $\tilde{\epsilon}_\text{max} \sim 0.01$ is more typical.

We emphasize that this upper bound on the fluctuation amplitude only applies for the analytical result presented in equation \ref{modR}; it does not apply to the numerical analysis of the perturbed differential equation (equation \ref{fluctBUDDENeq}). As figure \ref{BraggNUM} shows, the same qualitative behavior of the reflection coefficient in the small-amplitude regime persists to larger amplitudes, at least until additional SRs develop.

\section*{References}
\bibliography{/Users/Nick/Documents/Biblio.bib}
\bibliographystyle{iopart-num}

\end{document}